%% file: ceph-aph.tex
\documentstyle[11pt,aaspp4]{article}

\def\refitem{\relax}
\def\journref#1#2#3#4#5{\refitem #1 #2, #3, #4, #5}
\def\paspref#1#2#3#4{\journref{#1}{#2}{\pasp}{#3}{#4}}
\def\apjref#1#2#3#4{\journref{#1}{#2}{\apj}{#3}{#4}}
\def\apjsref#1#2#3#4{\journref{#1}{#2}{\apjs}{#3}{#4}}
\def\ajref#1#2#3#4{\journref{#1}{#2}{\aj}{#3}{#4}}
\def\mnrasref#1#2#3#4{\journref{#1}{#2}{\mnras}{#3}{#4}}

\def\kms{km~s$^{-1}$}
\def\etal{{\it et~al.}} 
\def\CORAVEL{{\sc coravel}}
\def\HI{H{\small I}}
\def\vminusi{\hbox{$V\!-\!I$}}               % V-I

\let\lta=\lesssim

\def\mytabcaption#1{\refstepcounter{table}{#1}\\}

\lefthead{Metzger et al.}
\righthead{Galactic Cepheid Kinematics}

\begin{document}

\title{The Shape and Scale of Galactic Rotation \\
	from Cepheid Kinematics}

\author{Mark R. Metzger\altaffilmark{1,2}}
\affil{Physics Department, Room 6-216, Massachusetts Institute of
Technology, \\
    Cambridge, MA 02139}

\author{John A.~R. Caldwell}
\affil{South African Astronomical Observatory, Observatory 7935, South Africa}

\and

\author{Paul L. Schechter}
\affil{Physics Department, Room 6-206, Massachusetts Institute of
Technology, \\
    Cambridge, MA 02139}

\altaffiltext{1}{Present Address:  California Institute of Technology,
Mail Code 105-24, Pasadena, CA 9115} 
\altaffiltext{2}{Email: {\tt mrm@grus.caltech.edu}}

\begin{abstract}

A catalog of Cepheid variables is used to probe the kinematics of the
Galactic disk.  Radial velocities are measured for eight distant Cepheids
toward $\ell = 300^\circ$; these new Cepheids provide a particularly good
constraint on the distance to the Galactic center, $R_0$.  We model the
disk with both an axisymmetric rotation curve and one with a weak elliptical
component, and find evidence for an ellipticity of $0.043 \pm 0.016$ near
the Sun.  Using these models, we derive $R_0 = 7.66 \pm 0.32$ kpc and 
$v_{\it circ} = 237 \pm 12$ \kms.  The distance to the Galactic center
agrees well with recent determinations from the distribution of RR Lyrae
variables, and disfavors most models with large ellipticities at the solar
orbit.

\end{abstract}


\section{Introduction}

The first use of Cepheids to measure kinematic parameters of the
rotation curve was by \nocite{joy39} Joy (1939), who found a distance to the Galactic
center of 10 kpc; this distance was inferred from the measured shape of the
rotation curve assuming a simple model. Cepheids have several significant
advantages over other stellar tracers for determining large-scale Galactic
kinematics.  They are intrinsically bright ($M_V \simeq -4.1$), which in
conjunction with variability makes them relatively easy to locate at large
distances.  Perhaps the most significant advantage of Cepheids is that
distances to them can be determined extremely well:  modern calibrations of
the period-luminosity relation in the near infrared yield uncertainties of
$< 5$\% (\cite{mf91}).  Distances to Cepheids in the Galactic
disk can be best obtained from near-infrared photometry, due both to the
smaller PL relation scatter and because of heavy extinction by dust in the
Galactic plane.

Further studies were carried out by Stibbs (1956),
Kraft \& Schmidt (1963), and Feast (1967), using additional data on Cepheids
and incorporating the use of individual reddenings in distance
measurements.  Their analysis was ultimately limited, however, by the small
amount of available data, particularly on distant, faint Cepheids.
\nocite{cc87} Caldwell \& Coulson (1987, hereafter CC)
made an extensive compilation of available Cepheid photometry and
radial velocities, and used the data in an axisymmetric rotation curve model
to determine, among other parameters, the distance to the Galactic center 
($R_0$).  Though a significant improvement over earlier work, their models
were also limited by the available data:  CC lamented that
many distant Cepheids lacked good radial velocities, and there were few
Cepheids known at large distances from the sun, particularly toward
directions which provide the best constraints on $R_0$.

More recently, new radial velocities for many distant Cepheids have
been measured (\cite{mcms91}, \cite{mcs92}, Pont \etal\ 1994b), and
used in models by Caldwell \etal\ (1992) and 
\nocite{pmb94} Pont, Mayor, \& Burki (1994a, hereafter PMB).
Uncertainties in the parameters of the axisymmetric rotation curve models
employed, including $R_0$, were significantly improved.  Under the
assumptions made by the models, it is one of the most precise ways to
measure $R_0$ (\cite{reid93}).

Many studies suggest, however, that the simple axisymmetric picture of
Galactic rotation may not be correct.  Much evidence leads to the conclusion
that the Milky Way is a barred spiral (e.g. \cite{bs91b}, \cite{bin91},
\cite{wein92}, \cite{dwek95}), and there are suggestions that even orbits near
the Sun are significantly elliptical (e.g. \cite{bs91a}, \cite{kt94}). A
curious result of many rotation models using Cepheids, first noted by
\nocite{joy39} Joy (1939), has been an apparent constant velocity offset 
of $\approx 3$ \kms\ between mean Cepheid velocities and the local standard
of rest (LSR).  Though the Cepheid sample has grown and measurements have
improved for recent studies, this term has persisted.  Some explanations for
this effect have included a possible error in radial velocity zero-point or
an artifact of measuring velocities in a pulsating atmosphere (CC). One of
the results of the PMB study was to suggest that instead this velocity
offset might be a kinematic effect explained by a bar-driven
ellipticity of the solar orbit.

In this paper, we present new data on eight distant Milky Way Cepheids
discovered by \nocite{cks91} Caldwell, Keane, \& Schechter (1991).  These
Cepheids are located near the solar circle, which gives them high leverage
in determining $R_0$ and helps to decouple the measurement of $R_0$ from the
local rotation speed.  We combine the new data with a catalog of Galactic
Cepheid data, and measure new Galactic rotation parameters using an
axisymmetric model.  We also study the kinematics using the non-axisymmetric
models of \nocite{kt94} Kuijken \& Tremaine (1994), and measure a component
of the local potential ellipticity.

\section{Velocities for New Cepheids}\label{sec:newcep}

\nocite{cks91} Caldwell, Keane, \& Schechter (1991) conducted a search for
distant Milky Way Cepheids in an area near $\ell = 300^\circ$, $b = 0^\circ$
covering 9.4 square degrees.  From over 2000 variable stars identified in
the survey, 37 were chosen as promising Cepheid candidates.  The candidates
were selected based on sparse I-band data, which could not provide a firm
classification of the candidates.  Additional multi-band photometry of the
candidates was obtained to help confirm the identity of these stars
(\cite{avr91}).  

We obtained spectra for many of the candidates in February 1991 at Las
Campanas, both to help establish these candidates as Cepheids and to obtain
radial velocities for use in kinematic models.  To select the most promising
candidates for frequent observation, we made use of V- and I-band followup
photometry of \nocite{avr91} Avruch (1991).  I-band data from the original
survey were combined with new observations to generate a detailed light
curve, and new V-band observations were used to provide color information at
different pulsational phases.  Since Cepheids have a characteristic
(\vminusi) color change over the course of pulsation (e.g. \cite{mb85}), we
ranked the candidates for observation based on the slope of $dV/dI$ as well
as the appearance of the I-band light curve.  

\subsection{Observations and Data Reduction}

Spectra of the candidates were taken with the Modular Spectrograph on the
DuPont 2.5m telescope at Las Campanas on the nights of 25 February
through 2 March 1991.  We used the spectrograph in a cross-dispersed mode
with a 150 $\ell/{\rm mm}$ immersion grating and a 300 $\ell/{\rm mm}$ grism
cross-disperser, projected onto a TI 800$\times$800 CCD.
The primary disperser was adjusted to place orders 14--25 onto the CCD,
providing coverage from 5000--8700 \AA.  A 1.0 arcsec slit was used
throughout the run, which projected to 2.2 pixels on the detector at 8400
\AA\ and gave an effective resolution of 60 \kms.  Calibration frames were
taken after each stellar spectrum using He-Ne and Fe-Ar lamps.

The data reduction was conducted using a slightly modified version of the
procedure described by \nocite{mcms91} Metzger \etal\ (1991).  Each spectrum
was flattened using an incandescent lamp exposure, and strong cosmic ray
events were removed.  Calibration lines from the associated lamp exposures
were identified and centroided using a modified version of DoPHOT
(\cite{doph89}), and fit across orders with a fifth-order 2-dimensional
Legendre polynomial.  The high-order coefficients were fixed using a long
lamp exposure that yielded over 300 identified lines, and the 4 lowest-order
coefficients were fit to each calibration frame, which typically had 100
available lines.  Each stellar spectrum was rebinned in log-$\lambda$
according to the calibration, and each order was separately extracted and
sky-subtracted.

\placefigure{fig:msflop}

During the course of the reduction, we noticed that the spectrum
shifted abruptly between two positions on the chip throughout the observing run.
The shift was aligned in the direction of the cross-dispersion, and it is
possible that the grism was not well secured, flopping between two
positions.  Figure \ref{fig:msflop} shows the position of the spectrum on
the chip as a function of telescope hour angle, from which it is clear that the
flop occurs near the meridian.  To account for the shift in the data
reduction, each spectrum and its associated calibration spectrum was
classified into a ``high'' or ``low'' group.  Separate flats and
high-order wavelength calibrations were made for each group and the
associated frames were reduced within its group.  Three frames that
suffered a shift during an exposure were discarded.

\placefigure{fig:mstemplvel}
\placetable{tab:l300std}

A velocity for each spectrum was calculated relative to a high
signal-to-noise spectrum of the star HD 83443 using the Fourier quotient technique of
\nocite{ssbs77} Sargent \etal\ (1977).  Several of the Cepheid candidates were too heavily
reddened to provide an adequate signal for radial velocities using
the blue orders, so we decided to use the
Ca triplet (8498, 8542, 8662 \AA) in order 14 to measure individual velocities.  
Several \CORAVEL\ faint southern radial velocity
standards (\cite{mau84}) were observed throughout the run, and were
used to calibrate the effective velocity of the template spectrum.  The
individual measurements of the standards (given as an effective
template velocity, computed using radial velocity of the standard star,
as by \cite{mcms91}) are shown in Figure~\ref{fig:mstemplvel}.  The
open and filled points correspond to ``low'' and ``high'' position spectra,
respectively.  The mean template velocities of the two groups are not
significantly different, confirming that the two groups of spectra have been
successfully referenced to the same zero point.  Table~\ref{tab:l300std} shows the mean
measured velocities for each radial velocity standard, along with velocities
for two metal-weak subdwarfs observed during the run.  Each velocity has
been adjusted upward by 0.4 \kms\ to bring these velocities to a minor
planet-based zero point (\cite{may85}; \cite{mcs92}, hereafter MCS).  Two error
estimates are given in Table~\ref{tab:l300std}: $\varepsilon_f$, the
formal error in the velocity from the Fourier quotient, and $\varepsilon_V
\equiv \sigma_v/\sqrt{n}$,
the standard error of the mean of the individual measurements.

\placetable{tab:l300vel}

The individual radial velocities for the stars confirmed as Cepheids are
given in Table \ref{tab:l300vel}.  The other variables did not have spectra
consistent with that of a Cepheid;  most of these did not have significant Ca
triplet absorption lines and thus did not yield radial velocities.

\placefigure{fig:l300gam}

\placetable{tab:l300gam}

\subsection{Gamma Velocities }\label{ssec:gamvel}

Gamma velocities were computed for each of the Cepheids according to the
method described by \nocite{mcs92} MCS.  For each star, the radial velocity
points are folded at the pulsation period and fit to a ``typical'' radial
velocity curve.  The shape of this velocity curve is fixed by the
photometrically determined period of the Cepheid, and is generated using
low-order Fourier coefficients that are functions of the period.  The
periods used for these stars were taken from \nocite{avr91} Avruch (1991),
and improved with additional data below.  The gamma velocity and phase are
then fit using a $\chi^2$ minimization procedure.

The radial velocities for each star and the curves fit to the points are
shown in Figure~\ref{fig:l300gam}.  Fit gamma velocities for the eight
Cepheids are given in Table~\ref{tab:l300gam}, along with the formal error
of the fit and the reduced $\chi^2$.  An additional error estimate, based on
a Monte Carlo simulation, is also given in Table~\ref{tab:l300gam}.  This
estimate takes the variation in the shape of the fit curve into account as
well as velocity measurement error.  One star thought to be a Cepheid from
its light curve (11582-6204), but suspect due to its near-infrared
photometry (\cite{sack92}), has a spectrum inconsistent with that of
a Cepheid and was discarded from the candidate list.

\subsection{Infrared Magnitudes}\label{ssec:irmags}

Mean $\langle K \rangle$ magnitudes of the new Cepheids were computed using
the data of \nocite{sack92} Schechter \etal\ (1992).  The values of
$\overline{K}$ they report are a straight average of their individual
measurements; if the measured points are not well-spaced in phase, such an
average can be biased with respect to the true $\langle K
\rangle$.  Though we expect this difference to be small given the pulsation
amplitude at $K$, to obtain a slightly more accurate average we fitted a
sine function to the $K$ points using periods computed above.  The amplitude
of the sine function was scaled from the $V$ light curve amplitude,
computed from the data of \nocite{avr91} Avruch (1991) and \nocite{led93}
LeDell (1993), 
using the
relation of \nocite{wel84} Welch \etal\ (1984):  ${\rm Amp}(K) = (0.30 \pm
0.03) \times {\rm Amp}(V)$.  The results of this procedure, along with
formal errors assuming $\sigma_K = 0.02$ mag for each observation, are given
in Table~\ref{tab:l300parms}.  Adding higher-order terms to the
light curves from the Fourier decompositions of \nocite{ls93a} Laney \&
Stobie (1993a) had no significant effect on the computed $\langle
K \rangle$.  The star 13240--6245 has a high covariance ($r \simeq 0.6$)
between $\langle K \rangle$ and the epoch, largely a result of poor phase
coverage.

Also shown in Table~\ref{tab:l300parms} are the amplitudes (peak-to-peak) of
the $V$-band light curves, epochs of maximum light in both $V$ and $K$, and
improved period estimates.  The $K$-band epochs were determined from the fit
light curve, and the improved periods are selected from values listed by Avruch
(1991) (he gives several due to the possibility of aliasing) that are most
consistent with the $K$ data.  We find that the $V$ maximum light lags that
in $K$ by $\sim 0.27$ cycles, in rough agreement with \nocite{wel84} Welch \etal\ (1984).
We also note that the star 13323--6224 is peculiar in that it has a significantly
smaller amplitude than expected given its period.  Overall, we obtain a
tight formal error on the $\langle K \rangle$ magnitudes; in particular, the
uncertainties are smaller than the scatter in the PL-K relation and hence
sufficient for our purposes.  (It is interesting to note that only two of
the average magnitudes are significantly different from the straight means
computed by Schechter \etal\ [1992].)  It was not necessary to phase and fit
the $(H-K)$ colors, since they do not change appreciably over the pulsation
cycle.

We used the $E(H-K)$ color excesses of Schechter \etal\ (1992) to compute the
extinction in $K$, $A_K \equiv K - K_0$, for the newly discovered Cepheids.
These were derived assuming an intrinsic color locus in the H--K/P plane:
$(H-K)_0 = 0.068 + 0.024(\log P - 1)$ (their equation 1).  We adopted the
same extinction law used by Schechter \etal\ for the total-to-selective
extinction, that given by \nocite{cfpe91}Cohen \etal\ (1981):  $A_K = 1.39 E(H-K)$.  This
can be compared with coefficients found in other sources: 1.7 (McGonegal
\etal\ 1983, CIT system), 1.5 (\cite{ccm89}, Johnson
system), 1.8 (\cite{rie85}), and 1.6 (Laney \& Stobie 1993a, Carter
system).  The Caldwell \etal\ Cepheids have an average $E(H-K)$ of 0.27; if we were to
simply replace our reddening law with the average of the $A_K/E(H-K)$ values
listed ($=1.6$), the result would be an increase in mean distance modulus by
0.06 mag for the stars in the sample.  This would not be strictly
correct, however, as the values are based on magnitudes of different systems; we use
the Cohen \etal\ (1981) value, keeping in mind a possible systematic offset.

\section{Galactic Cepheid Data}\label{sec:cepdat}

The new Cepheid data from \S\ref{sec:newcep} were combined with a catalog of
 Cepheid photometry and radial velocities
compiled from several sources.  We started with the
compilation of 184 stars by \nocite{cc87} CC.  Reddenings for many
additional Cepheids were obtained from \nocite{fer90}\nocite{fer94} Fernie
(1990, 1994), and new radial velocities from \nocite{mb87} Moffett \& Barnes (1987),
\nocite{mmb87} Mermilliod \etal\ (1987), and \nocite{mcs92} MCS.  Finally, mean B and V
photometry and many new radial velocities were obtained from the compilation
of \nocite{pmb94}PMB.  Known or suspected W Virginis stars (Pop. II
Cepheids), based on the list of \nocite{har85} Harris (1985), were excluded
from the sample.

\placefigure{fig:newred}

Because of the small scale height of Cepheids in the Galactic disk, the
distant stars lie within a few degrees of the Galactic plane in projection
and are thus significantly obscured by dust.  To measure
distances, accurate color excesses for the stars must be known to allow an
estimate of the total extinction.
Figure~\ref{fig:newred}a shows a comparison between the color excess given by
CC and the values of \nocite{fer90} Fernie (1990).  A clear trend is evident,
such that the redder stars tend to have higher values of E(B--V) on Fernie's
scale than that of CC.  A linear fit to the data gives
\begin{equation}
	E_{B-V}{\rm (CC)} - E_{B-V}{\rm (F90)} = 0.032 (\pm 0.006) - 0.107
		(\pm 0.013) \times E_{B-V}{\rm (F90)}
	\, ,
	\label{eq:fernieCCred}
\end{equation}

with a 0.05 mag scatter about the fit.  This trend implies a difference in
$R \equiv A_V/E(B-V)$ between the two systems of 10\%, or about 0.3 mag in
the corrected $V$ magnitude for the more heavily extinguished stars ($E_B-V
\simeq 1$).  The origin of the discrepancy is not clear, but it is
significant:  a 10\% change in $R$ corresponds to a 5\% change in $R_0$
derived from rotation curve models (see \S\ref{sec:rotmodels} below).  A
comparison between reddenings of Fernie (1990) and \nocite{dwc78} Dean,
Warren, \& Cousins (1978) for 94 stars in common, yields a similar but
shallower slope; Fernie shows a similar plot for cluster stars in his Figure
1.  Another comparison can be made to the recent study of Cepheid reddenings
by \nocite{ls94} Laney \& Stobie (1994), who combined optical and infrared
data to test the value of $R$.  Figure~\ref{fig:newred}b shows a comparison
of their reddenings to Fernie's, revealing the same trend as
Figure~\ref{fig:newred}a.  We have adopted the same value of $R$ as Laney \&
Stobie found consistent with their color excesses, thus to produce
extinction corrections we adjust Fernie's (1990) reddenings using
equation~\ref{eq:fernieCCred}.

\placetable{tab:cepdatA}

The combined data set contains 294 stars (not including the eight new stars
from \S\ref{sec:newcep}) and is shown in Table~\ref{tab:cepdatA}.  Values
reported for E(B--V) are primarily from Fernie (1990), transformed to our
system.

\subsection{Cepheid Distance Calibration}\label{ssec:cepcal}

Distances to the Cepheids were computed primarily via the
period-luminosity relation in the V-band (PL-V), which can be parameterized
\begin{equation}
	5 \log_{10} d =  m - m_0 - \alpha(\log P - 1) \, .
	\label{eq:perlum}
\end{equation}

Here $m_0$ corresponds to the unreddened apparent magnitude of a Cepheid
with a 10-day period at a distance of $R_0$, and $\alpha$ is the slope of
the adopted period-luminosity relation.  The measured period and unreddened
apparent magnitude for a particular Cepheid are given by $P$ and $m$,
respectively.  The stars were dereddened as in CC, using the prescription for
$R_V$ derived from Olson (1975) and Turner (1976):
\[
	R_V \equiv A_V/E(B-V) = 3.07 + 0.28 {(B-V)}_0 + 0.04 E(B-V) \, ,
	\label{eq:dered}
\]
whereby
\[
	m = \langle V \rangle - R_V\times E(B-V) \, .
\]
This extinction correction is also used by Laney \& Stobie (1993), and
has a slightly higher value for the effective $R_V$ than the reddening laws
of Savage \& Mathis (1979) and \nocite{rie85}Rieke \& Lebofsky (1985) (though the last two
do not give an explicit dependence on intrinsic color).  The dependence of
$R_V$ on the intrinsic color is caused by a shift in effective wavelength
of the filters when measuring stars of different spectral class (Olson 1975).
Extinction
corrections for infrared photometry are handled separately and
are described below.

We chose to avoid period-luminosity-color (PLC) relations and terms
accounting for metallicity in this study for two reasons.  First, the PL-V relation is
thought to be only weakly sensitive to metallicity, both in zero-point and
slope (Iben \& Renzini 1984, Freedman \& Madore 1990; but see also Caldwell
\& Coulson 1986), and less sensitive to metallicity than the
PLC relation (Stothers 1988).
Second, the intrinsic color used in the the PLC
relation is susceptible to significant error when this color is found by
dereddening using color excesses.  By making a correction based on color,
one must make some
implicit assumption about the intrinsic color of the star; any difference
between this assumption and the actual color (perhaps due to metallicity) is
amplified in the derived unreddened magnitude.
This process can
significantly increase the sensitivity of PLC distances to metallicity.  The
uncertainties in the reddening corrections themselves tend to be
larger than the effect of metallicity on the PL-V relation
over a wide range of metal abundance (Stothers 1988).    Further, the reddenings
derived for Cepheids themselves are sometimes derived under the assumption
of a unique color locus in the (B--V)/(V--I) plane (e.g. \cite{dwc78}).
Given these sources of error, and the
uncertainty in the slope of the PLC color term (Fernie \& McGonegal 1983,
Caldwell \& Coulson 1986), for this study we chose to use only PL relations (PL-V, plus the
PL-K relation for some of the models below) with no explicit correction for
a radial metallicity gradient.  The scatter in the individual distances may
be slightly higher, but the systematic errors are easier to quantify.

The two parameters $m_0$ and $\alpha$ in equation~\ref{eq:perlum} determine
the distances to each Cepheid in terms of $R_0$.  The PL-V slope parameter
$\alpha$ has been measured using both Magellanic Cloud and
Galactic cluster Cepheids (Fernie \& McGonegal
1983, Caldwell \& Coulson 1986, CC, 
Madore \& Freedman 1991, Laney \& Stobie 1994).  Most tend to agree to 
within the quoted errors, and lie in the
range $-2.9$ to $-2.8$ (with the exception of CC at $-3.1$).  
There does appear to be a significant difference in computed slope of the PL relation,
however,
depending on the period range of Cepheids used in the fit.  While studies
using open clusters to calibrate the PL relation contain data over a wide
range of period, many exclude the longest period Cepheids from the fit as
they tend to be somewhat brighter than an extrapolation of the PL relation
of short-period Cepheids would indicate (e.g. Fernie \& McGonegal 1983).
Freedman \etal\ 1993 derive a separate calibration of the PL
relation based only on Cepheids with $1.0 < \log P < 1.8$ to match most
closely the range of periods in the M81 Cepheids.  They find a PL-V slope of
$-3.35 \pm 0.22$, significantly steeper than when short-period Cepheids are
used.

A possible explanation for the discrepancy is the suggestion by
B\"ohm-Vitense (1994) that most Cepheids with periods shorter than 9 days
are overtone pulsators.  If the short and long period Cepheids form two
offset, steeper PL relations, then a slope measured from combining the two
would be shallower than that measured from either set independently.  More
work needs to be done to help verify the existence of the separate PL
relations, particularly in the near-infrared where the intrinsic scatter
about the PL relation is smaller.  A quick examination of the PL-K data of
Laney \& Stobie (1994) shows little evidence for short-period overtone
pulsators, while not necessarily ruling them out.  Gieren, Barnes, \&
Moffett (1989) find evidence against this hypothesis based on the continuity
of BW radii across a wide range of periods.  

There is direct evidence, however, that at least some of the Cepheids in our
sample are overtone pulsators.  For example, the recently measured radial
velocities for QZ Nor of $-38.6 \pm 0.7$ \kms\ (MCS) and V340 Nor of $-40.0
\pm 0.1$ \kms\ (\cite{mmb87}) confirm both as members of NGC 6067, and
support the conclusion of \nocite{cc85}Coulson \& Caldwell (1985) and 
\nocite{mb86}Moffett \& Barnes (1986) that QZ Nor is an overtone pulsator
while V340 Nor is not.  Given this conclusion, the period-luminosity
relation gives the same distance to both.  Other examples of such direct
evidence include SU Cas (\cite{eva91}).

Even with some contamination from overtone pulsators, so long as the range
of periods of the calibrators is similar to the overall population, the
derived slope and zero-point will still provide accurate distances (though
perhaps with larger scatter).  Considering that our Cepheid sample has a
median $\log P\approx 0.9$, we can comfortably use the shallower slopes
derived from Cepheids of similar period, and adopt a commensurate zero
point.  New data on Magellanic Cloud Cepheids discovered in the MACHO survey
(\cite{alv95}, \cite{alc95}) will help to answer this question conclusively.

The zero point of the Cepheid PL relation puts $m_0$, the unreddened
apparent magnitude of a Cepheid at a distance $R_0$, on an absolute distance
scale.  Various studies have yielded different Cepheid PL zero points, primarily
due to differences in assumed extinction, metallicity or correction for
metallicity, and the sample of stars used.  Currently the most accurate
methods for Galactic PL calibrations are those using Cepheids in clusters
and associations (\cite{tur85}; \cite{fm83}), and those using
the visual surface brightness (Baade-Wesselink) method (\cite{gie89}).  The
cluster calibrations are based on fitting main sequences for clusters
containing Cepheids to either the Hyades or Pleiades, and the surface
brightness method attempts to measure the radius of a Cepheid based upon
accurate photometry and radial velocity measurements.  A convenient
comparison of Cepheid calibrations can be made by applying the calibrations
to LMC Cepheids, and comparing the derived LMC distance moduli.  The SMC is
somewhat less suited to this purpose as it is thought to be significantly
extended along the line of sight.  \nocite{fw87} Feast \& Walker (1987) give a
comprehensive review of Cepheid calibrations up to that time, and conclude
that for a Pleiades modulus of 5.57, the LMC lies at a true distance modulus
of $18.47 \pm 0.15$.  This estimate is based on the same extinction law used
here.  More recently, using updated V-band data, CC determine an LMC modulus
of $18.45$, and Laney \& Stobie (1994) find $18.50 \pm 0.07$, both assuming
the same Pleiades modulus and extinction law.  The visual surface brightness
calibrations currently yield Cepheid distance moduli larger by $\sim 0.15$ mag on average
(\cite{gf93}), and give a distance modulus for the LMC of $18.71
\pm 0.10$ mag.  While significantly different, the Gieren \& Fouqu\'e data
appear to have an asymmetric distribution
that may make the distance moduli too large:  the four
calibrators they discard as being significantly discrepant all have distance
moduli too large by $> 0.6$ mag.  Though this does not completely resolve the
discrepancy, for this study we have chosen to adopt the cluster calibrations.
On this scale, our adopted V-band
calibration is 
\begin{equation}
	M_V = -4.10 - 2.87 (\log P - 1)
	\, .
	\label{eq:plvadopt}
\end{equation}
The
internal uncertainty in the zero point (exclusive of any systematic error
in the Pleiades/LMC distance) is estimated to be $\simeq 0.07$ mag.

To incorporate the near-infrared data on the newly-discovered Cepheids, we use the
period-luminosity relation in the $K$-band with an appropriate calibration
and extinction law.  The calibration zero point must give distances
commensurate with those derived from $V$-band data, and thus we again
normalize the zero-point to an LMC modulus of 18.50.  After making this
correction, the PL-$K$ calibration of Welch \etal\ (1987) gives $M_K = -5.66
- 3.37(\log P - 1)$, with the $K$ magnitudes on the same system (Elias
\etal\ 1982) as the Schechter \etal\ (1992) photometry.  Madore \& Freedman
(1991) give a self-consistent calibration based on a sample of 25 LMC Cepheids, each with
photometry in both V and K, finding $M_K = -5.70 - 3.42 (\log P - 1)$,
identical to within quoted errors.  The Madore \& Freedman PL-$V$ calibration
is also consistent with our adopted $M_V$.  Laney \& Stobie (1994)
give a calibration of the PL-$K$ relation in a slightly different
photometric system; after converting to the Elias \etal\ (1982) system using
the transformation of Laney \& Stobie (1993b), and adjusting to an LMC
modulus of 18.50, we find 
\begin{equation}
	M_K = -5.70(\pm0.04) - 3.40(\pm0.05) (\log P - 1)
	\, . \label{eq:plkadopt}
\end{equation}
The scatter of
the individual stars about the period-luminosity relation is significantly
smaller in $K$ than $V$, 0.16 mag rms vs. 0.25 mag rms, and hence the
internal error associated with the zero point is correspondingly smaller at
0.04 mag.  Since the quoted uncertainties in the Laney \& Stobie (1994)
calibration are the smallest of those quoted above, and that the relation is almost
identical to the others, we adopt equation~\ref{eq:plkadopt} for our models.

There is a small discrepancy between the LMC moduli derived from $V$ and $K$
data when using Galactic cluster calibrations.  Measurements of LMC distance
modulus in the $K$ band from the above references typically yield a value of
18.55--18.60, some 0.05--0.10 higher than the $V$ calibration.  This
discrepancy could be due to a number of factors, including a difference in
mean metallicity between Galactic and LMC Cepheids.  As the bolometric PL
relation for Cepheids is thought to be essentially independent of
metallicity (\cite{ir84}), the zero point of the PL-V and PL-K relations may
be expected to differ by $\sim 0.03$ mag over the metallicity difference of
the Galaxy and the LMC from the variation in bolometric correction alone
(\cite{ls94}, \cite{sto88}).  Another possible source of systematic error
arises from the correction for extinction: this is substantially larger for
the Galactic calibrators, which have a mean E(B--V) of 0.65 (\cite{fw87}),
than the LMC Cepheids, which have an E(B--V) of about 0.14 mag.  A
reasonable error of 0.1 in the adopted $R_V$ value would thus produce an
apparent distance offset between the two of 0.05 mag.  Since the difference
here is only slightly greater than 1 $\sigma$, no useful limits can be
placed on $R$ (or $A_V - A_K$).  However, we discuss below some implications
of the kinematic distance scale using the newly discovered Cepheids on the
adopted reddening law.

\section{Milky Way Rotation}\label{sec:rotmodels}

Figure~\ref{fig:rotschem} shows a geometric picture of a star with angular
velocity $\bf \Theta$ at a distance $D$ from the Sun, and at a distance $R_0$ from
the center of the galaxy.  The local standard of rest (LSR) rotates with
velocity $\bf \Theta_0$, and the Sun moves with a velocity peculiar to the LSR of
$(u_0,v_0,w_0)$ in the coordinates $(R,\Theta,Z)$.  For our models we assume
that the height above the disk, $|z| = D |\sin b|$, is sufficiently small
that the potential is dominated by the disk, and therefore a thin-disk model
adequately represents the orbits (i.e. the primarily rotational orbits of
the stars are decoupled from their vertical motion).  This condition is met
by the classical Cepheid population as they are confined to the disk with a
scale height of 70 pc (average absolute distance from the mean plane,\cite{ks63}).
We also fix the $Z$-component of the Sun's motion, $w_0$, at 7 \kms\
(\cite{del65}) as it is neither constrained by the data nor does it affect
the model.

The rotation models we employ take the form of a mean velocity field ${\bf
\Theta}(R,\phi)$ with components in the $\hat{R}$ and $\hat{\phi}$
directions.  The mean velocity of the stars
at $(R,\phi)$, as measured from the Sun, is 
\begin{eqnarray}
	v^{*}_r & = & \Theta_{R}(R_0,0)\cos\ell\cos b -
		\Theta_R(R,\phi)\cos(\phi+\ell)\cos b \nonumber\\
		&&   -\Theta_{\phi}(R_0,0)\sin\ell\cos b + \Theta_\phi(R,\phi)
			\sin(\phi+\ell)\cos b - v_{\odot *} \, .
	\label{eq:genrotmodel}
\end{eqnarray}
The Sun's peculiar motion relative to the LSR in the direction of
the star, $v_{\odot *}$, is given by
\begin{equation}
	v_{\odot *} = -u_0\cos\ell\cos b + v_0\sin\ell\cos b - w_0\sin b
	\, . \label{eq:pecvel}
\end{equation}

The data are fit to the models using a non-linear $\chi^2$ minimization
program.  Free parameters were determined by fitting the measured radial
velocities for each Cepheid to model velocities generated from the other
measured quantities ($\ell$, $b$, $\langle V \rangle$, $P$, and $E(B-V)$) via
equation~\ref{eq:genrotmodel}.  Each measured velocity was weighted as in CC
using the estimated radial velocity dispersion added in quadrature to the
effective velocity error introduced by the distance measurement:
\begin{equation}
	\sigma^2_i = \sigma^2_v + \sigma^2_d {(\partial v_r / \partial d)}^2
\, . \label{eq:velsigma}
\end{equation}
The dispersion in the radial velocities is a combination of measurement
error and the intrinsic velocity dispersion of the stars in the disk (the
latter dominating), and was taken to be $\sigma_v = 11$ \kms.  The random
error in the distance from all sources (measurement, extinction correction,
and PL dispersion), $\sigma_d$, was assumed to be 0.2 mag.

\subsection{Axisymmetric Models}\label{ssec:aximodels}

The initial models we use to derive parameters of Galactic rotation are
based on a linear, axisymmetric rotation curve in a manner similar to
\nocite{cc87} CC.
An axisymmetric
rotation curve is given by $\Theta_R = 0$ and $\Theta_\phi(R)$ independent
of $\phi$.  If we make the approximation that the rotation curve is linear,
i.e. that
\[
	\Theta_\phi(R) \, \approx \, \Theta_0 + (R-R_0) {\left( \frac{d\Theta}{dR}
		    \right)}_{R_0} \!\!
                  = \, \Theta_0 + (r-1) {\left( \frac{d\Theta}{dr}
			\right)}_{r=1},
\]
equation~\ref{eq:genrotmodel} reduces to (cf. \cite{mb81}, CC)
\begin{equation}
	v^*_r = -2AR_0 \left( 1-\frac{1}{r}\right) \sin\ell\cos b - v_{\odot *}\, .
	\label{eq:rvrot}
\end{equation}
Here we have defined $r \equiv R/R_0$, and A is Oort's constant
\[
	A \equiv \frac{1}{2} \left[ \frac{\Theta_0}{R_0} -
		 {\left(\frac{d\Theta}{dR}\right)}_{R_0} \right] \, .
\]
The transformation of the heliocentric distance $d \equiv D/R_0$
to the Galactocentric distance $r$ is given by
\begin{equation}
	r^2 = 1 + d^2 - 2d\cos\ell \, .
	\label{eq:dtor}
\end{equation}
While the expansion of the rotation curve is to first order in $r$, 
and thus is valid primarily for stars near the solar circle,
no approximation is made for small $d$.

An additional parameter was added to
compensate for a possible zero-point offset in the radial velocities 
($\delta v_r$).  The full
model for the measured heliocentric radial velocity of a star is then
\begin{eqnarray}
	v_r & = & -2AR_0 \left( 1-\frac{1}{r}\right) \sin\ell\cos b \nonumber \\
		&& {}+ u_0 \cos\ell\cos b \nonumber\\
		&& {}- v_0 \sin\ell\cos b \nonumber\\
		&& {}- w_0 \sin b - \delta v_r \, .
	\label{eq:rotmodel}
\end{eqnarray}
Note that the sign of $u_0$ follows a Galactic radial convention and is opposite
that used by CC.
The model has parameters $2AR_0$, $m_0$,
$\alpha$, $u_0$, $v_0$, $w_0$, and $\delta v_r$; $\alpha$ was fixed according
to the PL relation adopted, $w_0$ was fixed as described above, and the
remaining parameters were fit using the $\chi^2$ minimization routine.

\placetable{tab:modelparA}

Table~\ref{tab:modelparA} gives a summary of the results of the axisymmetric
models.  Model A1 includes all of the sample Cepheids, including the 8 new
Cepheids with distances determined from K-band photometry and reddenings.
Distances to the remaining stars were computed using the color excess scale
given by equation~\ref{eq:fernieCCred}.  Model A1.1 is identical except for
the exclusion of 6 stars with large residuals in model A1: FF Car, FM Car,
BB Gem, VW Pup, AA Ser, and V Vel.  The model parameters remain essentially
the same, while the associated uncertainties fall (the six excluded stars
contribute almost 25\% of the residual error in model A1).  The error on
$R_0$ and $2AR_0$ are independently less than 5\%; the high covariance of
the two errors (see below) implies an even tighter constraint on $A =
2AR_0/2R_0$ $= 15.4 \pm 0.3$ km s$^{-1}$ kpc$^{-1}$.  It is important to note
that the errors given are internal errors (those due to scatter about the
adjustable parameters), and do not include systematic uncertainties
associated with the fixed model.  Model A1.2 uses the same stars as A1.1,
but employs the color excesses of Fernie (1990) directly, without adjustment.
The most significant effect is a shortening of the distance scale by $\sim 5$\%.

As a check, we ran our modeling software on the data used by CC in their
study.  The parameters are listed as model A2 in Table~\ref{tab:modelparA},
and agree well with those determined by CC.  The largest difference between
the A1 and A2 parameters, aside from the reduced uncertainties, is the value
determined for $2AR_0$.  Model A2.3 was run on the same set of Cepheids, but
with updated velocities from Table~\ref{tab:cepdatA} and reddenings from
Fernie (1990).  Again, $m_0$ decreases by about 5\%, while $2AR_0$ increases
by $0.8\sigma$.  Oort's $A$ constant effectively increases from
14.2 km s$^{-1}$ kpc$^{-1}$ to 15.7 km s$^{-1}$ kpc$^{-1}$ when the new data
are used.  Models A2.2 and A2.3 show the effect of adding the new velocities
and reddenings separately.

We also fit our models to the data set used by PMB as a further test.  Model A3
uses their pruned set of 266 stars, and further the parameter $\delta v_r$
was fixed at zero as in their study.  Our model reproduces their results
closely, though our estimates of uncertainty in the model parameters are
somewhat larger than reported by PMB. 

Since the PMB study employed color excesses on the F90 scale, the effective
distance scale should be the same between model A3 and A1.2.  However, the
computed $R_0$ is somewhat larger in model A3 than A1.2, indicating a
difference in the implied distance to the Galactic center.  At first we
guessed this might be due to the effect of adding the eight new distant
Cepheids to the model--the high leverage on $R_0$ for these stars could
produce the change (particularly since distances were computed using a
separate PL relation).  Deleting these stars, however, only increased the
uncertainty of $R_0$ without changing the most likely value.  We found that
the discrepancy was caused by including the $\delta v_r$ parameter in the
model:  the values determined for $R_0$ and $2AR_0$ fall by 1-$\sigma$
(model A3.1 vs. model A3), and resolve the discrepancy.  We conclude that
not including the $\delta v_r$ in the axisymmetric models skews PMB's $R_0$
measurement slightly high.  Allowance for this term can be made in
non-axisymmetric models, as PMB suggest, and is treated in
\S\ref{ssec:nonaxi}.

\placetable{tab:modelparA1}
\placefigure{fig:chicontour}

We can further determine the effect of the new Cepheid data by examining the
covariances between model parameters.  Table~\ref{tab:modelparA1} shows the
covariances of the model parameters determined using different sets of
data.  Covariances are expressed here as correlation coefficients of the
projected data \nocite{bev69}(Bevington 1969):
\[
	r_{ij} = \frac{s^2_{ij}}{s_{ii} s_{jj}}
\, 
\]
where the $s^2_{ij}$ are elements of the covariance matrix.  The
reduction in the covariance between $m_0$ and $2AR_0$ in the model including
the new Cepheids (A1.1) over that without (A1.4) is due to the advantageous
placement of the new Cepheids.  Because they lie at a Galactocentric radius
near the solar circle, the mean radial velocity of the stars in the model is
close to zero, independent of rotation speed.  (In an axisymmetric model,
stars lying at the same radius rotate together in a ring with no relative
velocity, save for random motion.)  Thus adding only a few stars serves to
decouple the two parameters in the model.  Figure~\ref{fig:chicontour} shows
a comparison between constant $\chi^2$ contours in $m_0$, $2AR_0$ space with
and without the eight stars.

The eight new CKS Cepheids turn out to provide a useful constraint on $R_0$
by themselves, assuming axisymmetric rotation.  Fitting the eight Cepheids
to the model, we derive $R_0 = 8.1 \pm 0.5$ kpc, assuming $v_0 = 14$ km
s$^{-1}$.  Thus with just eight stars, knowing $v_0$ gives us a distance to
the Galactic center to 6\% precision.  If we remove the nearest Cepheid,
13323-6224 at $0.3 R_0$, most of the remaining covariance between $R_0$ and
$2AR_0$ is eliminated.  The error reduces to $\sim 4$\%, with $R_0 = 7.95
\pm 0.31$ kpc.  The parameters $R_0$,$v_0$ have a high covariance in this
model; decreasing $v_0$ by 4 \kms\ reduces the derived value of $R_0$ by
5\%.  Again, all of these models rely on the assumption of an axisymmetric
rotation curve; in the next section, we relax this assumption.

\subsection{Non-Axisymmetric Models}\label{ssec:nonaxi}

The persistence of a significant $\delta v_r$ in the Cepheid data, even
after significant improvements in radial velocities and distances, leads to
the conclusion that either some systematic error is present in measuring
$\gamma-$velocities of Cepheids, or that an axisymmetric model is not
sufficient to describe the rotation curve.  As many Cepheids in open
clusters now have accurate $\gamma$-velocities within 0.4 \kms\ of the
cluster mean velocity (e.g. \cite{mmb87}), the systematic error in measuring
$\gamma$ is probably small.  Based on a comparison with an N-body
simulation, PMB suggest that the effect could be due to non-axisymmetric
motion driven by a central bar of $\sim 5$ kpc in extent.  Here we examine
the shape of the local rotation ellipticity directly using a simple
non-axisymmetric model.

Our non-axisymmetric models are based on those of \nocite{kt94} Kuijken \&
Tremaine (1994, hereafter KT).  The rotation curve is produced by a
primarily axisymmetric potential with a small $m=2$ perturbation, with minor
axis in the direction $\phi_b$.  The circular velocity and potential ellipticity
have a power-law dependence on radius:
\begin{equation}
	v_c(R) = \Theta_0
	{\left(\frac{R}{R_0}\right)}^{\alpha} \; ; \;\;\;\;
	\epsilon(R) = \epsilon_0 {\left(\frac{R}{R_0}\right)}^{p-2\alpha}\! .
\end{equation}
The rotation curve is given by equation~(\ref{eq:genrotmodel}) and by
(KT, eqn. 5a):
\begin{eqnarray}
	\Theta_R(R,\phi) & = & \beta_1 v_c(R) [s(R)\cos 2\phi
		- c(R)\sin 2\phi] \, , \nonumber\\
	\Theta_\phi(R,\phi) & = & v_c(R) - 
		\beta_2 v_c(R) [c(R)\cos 2\phi - s(R)\sin
		2\phi] \, ,
	\label{eq:nonaximodel}
\end{eqnarray}
where we have defined
\begin{equation}
	\beta_1 \equiv \left(\frac{1+p/2}{1-\alpha}\right)\, , \;\;\;\;
	\beta_2 \equiv \left(\frac{1+p(1+\alpha)/4}{1-\alpha}\right)
\end{equation}
and two orthogonal projections of the ellipticity
\begin{equation}
	c(R) \equiv \epsilon(R)\cos 2\phi_b\, , \;\;\;\;
	s(R) \equiv \epsilon(R)\sin 2\phi_b \, .
\end{equation}
The $c(R)$ and $s(R)$ parameters correspond to components of the ellipticity
that are symmetric and antisymmetric, respectively, about the Sun-center
line $\phi=0$.

As pointed out by KT, if the Sun lies near a symmetry axis of a
non-axisymmetric distortion, it is not easily detected using data such as
our sample where $d\lta R_0$.  In particular, the parameters $c$ and $R_0$
in the models have a high covariance, and thus do not produce independent
information.  If such a distortion were present ($c$ non-zero), the derived
parameters $R_0$ and $2AR_0$ can deviate significantly from their true
values.  To judge the size of the effect for our data, we have fit a simple
model with a flat rotation curve and constant ellipticity ($p = \alpha = 0$),
and allow the symmetric ellipticity component $c$ to vary along with the
other parameters.  For this model, the predicted mean radial velocities
reduce to
\begin{eqnarray}
	v_r & = & \Theta_0 \left( \frac{1}{r}-1\right) \sin\ell\cos b + \nonumber \\
		&& + \frac{\Theta_0}{r} c(2d\cos\ell-1) \sin\ell - v_{\odot *}\, .
\end{eqnarray}

\placefigure{fig:kt1}

As expected, if we fit for all six parameters, we find a covariance between
$2AR_0$, $m_0$, and $c$ of 0.95---too large to make any independent
constraints on these parameters.  We can, however, fix the symmetric
ellipticity in the models and determine how this affects the other derived
parameters.  Figure~\ref{fig:kt1} shows how $m_0$ and $2AR_0$ vary as a
function of the ellipticity parameter.  From a combination of non-kinematic
estimates of $R_0$ (\cite{reid93}), we can deduce a weak constraint on the
ellipticity at the solar circle $-0.08 < c(R_0) < 0.14$.

\placetable{tab:modelparB}

Because $c$ is degenerate with other parameters, we fix it at zero
and investigate the antisymmetric term $s$.  For these models we incorporate
the same data set used in model A1.1, and for the primary model assume a
flat rotation curve and constant ellipticity ($\alpha = p = 0$).  Here we
also assume that the velocity offset $\delta v_r$ is a kinematic effect of
the other model parameters, and therefore fix $\delta v_r = 0$.
Table~\ref{tab:modelparB} shows the results of several models.  Model B1
gives results for the primary elliptical model, and shows a $2.5 \sigma$
detection of ellipticity, $s(R_0) = 0.043 \pm 0.016$.  The B2 models vary the
extinction coefficient $R$, and B3 models vary $p$ and $\alpha$; the
ellipticity detection is robust in all except the $p>0$ models.  Note that
while the implied circular velocity increases rapidly as $\alpha$ increases,
Oort's A remains roughly constant.  Velocity residuals for model B1 are
shown in Figure~\ref{fig:b1resid}; there is some hint that systematic
velocity structure not accounted for by the simple models is present,
perhaps due to the influence of spiral structure.

\section{Discussion}\label{sec:discuss}

The uncertainties quoted in Tables~\ref{tab:modelparA}--\ref{tab:modelparB}
correspond to internal error estimates in the models, derived from the
scatter of the data about the fit model.  As described in
\S\ref{ssec:cepcal} above, systematic uncertainties in the distance scale
are also present.  To summarize, the internal uncertainty in the $V$-band
zero point is $\simeq 0.07$ mag, to which we add an estimated $0.1$ mag
error in the Pleiades modulus; similarly, the internal uncertainty in the
$K$-band calibration is $0.04$ mag.  Thus we adopt a calibration error of
$0.12$ mag and $0.11$ mag for $V$ and $K$ zero points, respectively.  There
is also an uncertainty associated with the reddening scale for the $V$-band
distances; however, we note that since we have a good measurement of $R_0$
based only on $K$-band photometry, this uncertainty does not contribute to
the overall distance scale error.  The reddening uncertainty in the $K$-band
corresponds to $\sim 0.05$ mag in $R_0$.  The $K$-band distance scale agrees
closely with that derived from the full models using the Dean \etal\ (1978)
optical reddening scale, thus we adopt that scale for our derived
distances.  The largest systematic uncertainty in measuring $R_0$, however,
derives from the inability of the rotation models to adequately constrain
$c(R_0)$.  KT suggest, based on a combination of local kinematics and global
\HI\ distribution, that $c(R_0) = 0.082 \pm 0.014$.  This corresponds to
overestimating $R_0$ in the kinematic models by $\sim 15$\% (see
Figure~\ref{fig:kt1}).  For now, we keep this component of the systematic
uncertainty separate; below we see that based on the RR Lyrae distance
scale, $c(R_0)$ is likely small.  Our standard model B1 then gives $R_0 =
7.66 \pm 0.32 \pm 0.44$ kpc and $v_c(R_0) = 237 \pm 12 \pm 13$ \kms.
Computing Oort's A from these gives $A = 15.5 \pm 0.4 \pm 1.2$ \kms\
kpc$^{-1}$.

Since the models of CC, the number of Cepheids with data adequate for use in
rotation models has increased by 50\%, in particular the addition of stars
at large distance from the Sun that provide good leverage for measuring
scale and shape parameters.  The internal uncertainty in $R_0$ and $2AR_0$
has decreased by approximately a factor of two, while remaining consistent
in their mean values.  In comparison with the recent PMB models, our
measurement of $R_0$ is slightly smaller than that found by PMB, 8.1 kpc;
much of this difference can be attributed to the absence of a direct
ellipticity term in the PMB models.  This also appears to skew the
measurement of $2AR_0$ to the high side; our derived values differ by almost
2-$\sigma$.  We find that other differences between our models, including a
higher-order expansion of the rotation speed and a different minimization
algorithm, produce negligible differences in the fit parameters.

Our results compare favorably with other methods of determining $R_0$.
Reid (1993) computes a ``best value'' for $R_0$, based on a weighted average
of many techniques, of $8.0 \pm 0.5$ kpc, in agreement with the results from
this paper.  In particular, Reid divides measurements into several
categories; our present results fall into two different categories, one that
is based primarily on the calibration of the $K$-band Cepheid PL relation,
and one that is based on a kinematic rotation model.  Our distance is
somewhat larger than that implied by a direct measurement of the distance to
Sgr B2 at the Galactic center of $7.1 \pm 1.5$ kpc (\cite{reid88}), though
still well within the errors.  A direct measurement of the proper motion of
Sgr A$^*$ by \nocite{bs87} Backer \& Sramek (1987), when interpreted with our measured
circular velocity, translates to $8.3 \pm 1.0$ kpc, again consistent with
our measurement, though the uncertainty is a factor of 2--3 larger.  

We can also use $R_0$ as a standard length to compare the Cepheid and RR
Lyrae distance scales.  Based on new near-infrared photometry of RR Lyraes in
Baade's window, \nocite{car95} Carney \etal\ (1995) measure $R_0$ to a
precision of about 5\%.  This result is much less sensitive the uncertain
extinction correction than the previous optical work (e.g. \cite{wlt91}),
and is also independent of the rotation curve shape.  Carney \etal\ quote
two different values for $R_0$: $7.8 \pm 0.4$ kpc, based on the zero-point
calibration of Carney, Storm, \& Jones (1992, CSJ); and $8.9 \pm 0.5$ kpc,
based on an RR Lyrae calibration from the LMC corresponding to an LMC modulus
of 18.50.  As noted by Walker (1992), the Cepheid LMC distance modulus is
0.2--0.3 mag higher than that derived from LMC RR Lyraes on the CSJ zero
point.  Our result for $R_0$ agrees quite well with the CSJ-scale distance,
and is about 2-$\sigma$ below the LMC RR Lyrae-based calibration.  The
systematic change in our $R_0$ measurement implied by increasing $c(R_0)$ is
in the opposite sense to that implied by adopting an LMC RR Lyrae distance
modulus of 18.50:  if the KT suggestion that $c(R_0)
\simeq 0.08$ is correct, it would only make our estimate for $R_0$ shorter
and the discrepancy with the LMC RR Lyrae calibration larger.  Thus the CSJ
calibration of RR Lyraes and the Cepheid distance scale appear to be
commensurate to within $\sim 0.15$ mag, and the ellipticity term $c(R_0) <
0.04$.  This implies that a small but probably real discrepancy exists
between Cepheid and RR Lyrae LMC distances, and that the intrinsic
magnitudes of the populations in the Galaxy and LMC may be different
(\cite{vdb95}; \cite{gou94})..  The two distance scales appear to be in
agreement, however, when Cepheids are compared with globular cluster
horizontal branch magnitudes in M31 (\cite{ajh96}).  

\placefigure{fig:ellipschem}

A positive value of $s(R_0)$ such as we find indicates that the Sun's orbit
is elliptical with major axis in the quadrant $0 < \phi < \pi/2$,
illustrated in Figure~\ref{fig:ellipschem} (see also Schechter 1996).  PMB
suggest that the orbital ellipticity could be due to a bar of roughly 5 kpc
radius, based on results of a numerical simulation; this is somewhat larger
than the $\lta 3$ kpc scale suggested by photometric observations (e.g.
\cite{wein92}).  The kinematic signature of the smaller-scale bar at $R_0$ is
predicted to be $\lta 1$\% in the models of
\nocite{wein94} Weinberg (1994), thus would be a strict lower limit for the
scale of the bar if the outward motion of the LSR is such a kinematic
signature.  The B3 models in Table~\ref{tab:modelparB} with $p<0$ correspond to
these inner bar models with ellipticity falling at $R_0$.

Our measurement of $s(R_0)$ indicates that the Sun has a small outward
velocity component with respect to the GC, of $\Theta_R(R_0,0) = 10.2 \pm
3.8$ \kms\ for our fiducial model B1.  This model has $p=0$, i.e. the
ellipticity is constant with radius, which predicts that stars toward $\ell
= 180^\circ$ should have zero mean velocity.  Observations of old disk stars
at the anticenter, however, indicate a mean motion of $\sim 6$ \kms in the
sense that the LSR and the outer Galaxy stars are moving {\it apart}
(\cite{lf89}, \cite{ms94}).  This would suggest that $s(R)$ increases with
radius, such as might be the effect of a triaxial halo that gradually
dominates an axisymmetric disk potential as $R$ increases.  Orbits at larger
radius would thus have larger ellipticity, implying a larger $<v_r>$ (so
long as $\alpha
\ge 0$), and would thus appear to have a net outward motion when viewed from
inner orbits.  This corresponds to models B3 with $p>0$; when we fit to
these models, however, the magnitude of local ellipticity shrinks and
becomes negligible with $p = 1$.
The outward motion is consistent with the 14 \kms\ value of the
\nocite{bs91a} Blitz \& Spergel (1991a) model derived from outer Galaxy gas
motions; however, warps and possible lopsided structure of the outer gas
make the velocity structure difficult to interpret (\cite{kui92}).

One of the obstacles to drawing strong conclusions from models with
different $p$ and $\alpha$ is that the distribution of the Cepheids in the
model is both asymmetric with respect to the GC and concentrated about $R=R_0$.
Constraining $s(R)$ is best accomplished when the antisymmetric component
can be measured directly on {\it both} sides of the Galactic center.  To
help fill out the region $0 < \ell < 90^\circ$, a survey for distant
Cepheids has been conducted toward $\ell = 60^\circ$ with some success
(\cite{met94}).  Further surveys are in progress to find Cepheids
at even greater distances on the opposite side of the galaxy; if a sample of
such Cepheids were available with radial velocities, good constraints could
also be placed on $c(R)$ (KT).

It is clear that given the uncertainties in estimating total $V$-band
extinction toward heavily obscured Cepheids (\S\ref{sec:cepdat}), the future
direction of Cepheid kinematic models will be toward using infrared
photometry.  This will be particularly important for new samples of distant
stars; as an example, some of the newly-discovered Cepheids toward $\ell =
60^\circ$ have E(B-V) of $\sim 4$ mag.  An uncertainty in $R_V$
of only 0.1 corresponds to a minimum distance error of 20\%, while in $K$
the corresponding extinction and error is reduced by a factor of 10.
Future surveys for
distant Cepheids, particularly those to find Cepheids at distances $>R_0$,
are also best conducted in the infrared.  The small scale height of Cepheids
is of order that of the gas and dust in the disk, so the total extinction to
such stars on the opposite side of the Galaxy is prohibitive in the optical.

\section{Summary}\label{sec:summary}

We have obtained new $\gamma$-velocities for the eight Cepheids discovered
by CKS, and combine these with $K$-band photometry of \nocite{sack92}
Schechter \etal\ 1992 for use in disk kinematic models.  Using a catalog of
Cepheid magnitudes, color excesses, and radial velocities collected from many
sources, we examine Galactic rotation parameters in the context of both
axisymmetric and non-axisymmetric models.  In our adopted model, which
assumes a flat rotation curve and constant ellipticity near the sun, we find
\begin{eqnarray}
	&R_0 = 7.66 \pm 0.32 \pm 0.44 \; {\rm kpc} \, , \nonumber \\
	&v_c(R_0) = 237 \pm 12 \pm 13 \; {\rm km \; s}^{-1} \, , \nonumber \\
	&s(R_0) = 0.043 \pm 0.016 \, , \nonumber \\
	&A = 15.5 \pm 0.4 \pm 1.2 \; {\rm km \; s}^{-1} \; {\rm kpc}^{-1} \, . \nonumber
\end{eqnarray}
The errors quoted are internal and systematic, respectively.  An additional
systematic error is also present from the unknown component of ellipticity
that is symmetric about the sun-center line, but comparisons with recent
measurements of $R_0$ from RR Lyraes indicate that this component should be
small.  An estimate of $R_0$ can also be made directly from 7 Cepheids near
the solar orbit with infrared photometry, independent of the rotation speed,
giving $7.95 \pm 0.31$ kpc for zero ellipticity and $7.61 \pm 0.30$ kpc for
4\% ellipticity.  There is good agreement between the new values and
some previously published estimates, with significantly smaller uncertainty in
the present results.  Future progress in determining rotation curve scale
and shape will rely on infrared searches and photometry, both to find more
distant stars through heavy extinction and to measure the distances more
accurately.

\acknowledgments

We thank the Director of the Carnegie Observatories, L. Searle, for use of
the facilities at Las Campanas, and the assistance of F. Peralta and O.
Duhalde at the telescope.  This work was supported by NSF grants
AST-8996139, AST-9015920, and AST-9215736.

\appendix

\clearpage

\clearpage

\clearpage
 
\begin{table}
	\centering
	\mytabcaption{Table~\ref{tab:l300std}. Radial Velocity Standards, Feb 91}
		\label{tab:l300std}
	\vspace{2mm}
\newdimen\digitwidth
\newdimen\minuswidth
\setbox0=\hbox{\rm0}
\digitwidth=\wd0
\catcode`?=\active
\def?{\kern\digitwidth}
\setbox0=\hbox{\rm-}
\minuswidth=\wd0
\catcode`!=\active
\def!{\kern\minuswidth}
{ \renewcommand{\baselinestretch}{1.0} \tabcolsep=0.4cm
\small\normalsize % this should ensure baselineskip gets adjusted
\begin{tabular}{lcccr}
\hline
\hline
??Star  & ?!$V_r$ & $\varepsilon_f$     & $\varepsilon_V$       & $n$ \\
\hline
HD 24331	& ?!26.6	& 1.7  & 1.3	& 3  \\
HD 39194	& ?!16.4	& 1.7  & 0.8	& 4  \\
HD 48381	& ?!40.1	& 1.5  & 0.5	& 4  \\
HD 83443	& ?!27.4	& 1.2  & 0.7	& 6  \\
HD 83516	& ?!43.3	& 1.2  & 1.5	& 7  \\
HD 101266	& ?!21.5 	& 1.3  & 1.5	& 6  \\
HD 111417	& ?-18.9	& 1.2  & 0.8	& 7  \\
HD 176047	& ?-41.5	& 1.3  & 0.6	& 5  \\
CPD -43$^\circ$ 2527	
		& ?!19.3	& 1.3  & 1.2	& 5  \\
                &               &      &        &    \\
HD 74000        & !205.5        & 4.6  & 1.6    & 4  \\
HD 140283       & -171.2        & 3.2  & 3.6    & 3  \\
\hline
\hline
\end{tabular}
} % tabcolsep redef

\end{table}

\begin{table}
	\centering
	\mytabcaption{Table~\ref{tab:l300vel}. Cepheid Radial Velocities, Feb 91}\label{tab:l300vel}
	\vspace{2mm}
\newdimen\digitwidth
\newdimen\minuswidth
\setbox0=\hbox{\rm0}
\digitwidth=\wd0
\catcode`?=\active
\def?{\kern\digitwidth}
\setbox0=\hbox{\rm-}
\minuswidth=\wd0
\catcode`!=\active
\def!{\kern\minuswidth}
{ %\tabcolsep=0.4cm
\renewcommand{\baselinestretch}{1.0} \small\normalsize % force adjust
\begin{tabular}{lrr|lrr}
\hline
\hline
\multicolumn{1}{c}{JD} & \multicolumn{1}{c}{V$_r$} &
\multicolumn{1}{c}{$\sigma$} & \multicolumn{1}{|c}{JD} &
\multicolumn{1}{c}{V$_r$} & \multicolumn{1}{c}{$\sigma$} \\
-2400000 & km s$^{-1}$ & & -2400000 & km s$^{-1}$ & \\
\hline
 11447-6153 &&                & 12003-6213 &&                \\
 48312.748 &    23.8  &  4.0  & 48312.730 &    -3.9  &  4.1  \\
 48313.649 &    31.0  &  3.6  & 48313.754 &   -12.3  &  3.8  \\
 48314.661 &    42.8  &  3.2  & 48314.770 &    -3.5  &  3.4  \\
 48315.651 &    52.0  &  4.3  & 48315.728 &    11.8  &  4.1  \\
 48316.726 &    11.3  &  5.3  & 48316.806 &    24.5  &  3.4  \\
 48317.612 &    13.9  &  5.2  & 48317.709 &    20.0  &  4.6  \\
 &&                           & &&                           \\
 11465-6209 &&                & 13190-6235 &&                \\
 48312.679 &   -12.4  &  5.7  & 48312.794 &   -35.4  &  3.8  \\
 48313.581 &   -13.9  &  4.6  & 48313.812 &   -22.6  &  4.0  \\
 48314.598 &    -2.7  &  3.0  & 48314.820 &   -14.2  &  3.1  \\
 48315.606 &   -15.4  &  4.4  & 48315.817 &    -8.5  &  4.3  \\
 48316.676 &   -18.5  &  4.0  & 48316.877 &   -24.5  &  4.3  \\
 48317.577 &    -0.4  &  3.5  & 48317.770 &   -45.4  &  4.7  \\
 &&                           & &&                           \\
 11492-6257 &&                & 13240-6245 &&                \\
 48312.847 &    10.2  &  7.3  & 48313.765 &    -7.5  &  3.7  \\
 48313.682 &    -6.1  &  4.3  & 48314.831 &    -9.4  &  3.8  \\
 48314.742 &     4.8  &  5.6  & 48315.880 &   -26.6  &  5.0  \\
 48315.712 &    20.0  &  6.6  & 48316.830 &   -32.2  &  5.7  \\
 48316.757 &    -6.7  &  5.2  & &&                           \\
 48317.676 &    -3.1  &  4.6  & &&                           \\
 &&                           & &&                           \\
 11521-6200 &&                & 13323-6224 &&                \\
 48312.693 &    31.4  &  4.4  & 48313.782 &   -39.7  &  4.8  \\
 48313.596 &     4.4  &  3.9  & 48314.851 &   -54.9  &  4.9  \\
 48314.695 &    12.6  &  3.6  & 48315.869 &   -53.3  &  4.5  \\
 48315.620 &    14.1  &  4.9  &	48316.847 &   -41.9  &  3.7  \\
 48316.701 &    19.9  &  3.1  &	48317.799 &   -37.2  &  4.5  \\
 48317.638 &    26.0  &  3.9  & &&                           \\
\hline
\hline
\end{tabular}
} % tabcolsep redef

\end{table}

\begin{table}[p]
	\centering
	\mytabcaption{Table~\ref{tab:l300gam}. Gamma Velocities of New Cepheids}\label{tab:l300gam}
	\vspace{2mm}
\newdimen\digitwidth
\newdimen\minuswidth
\newdimen\colonwidth
\setbox0=\hbox{\rm0}
\digitwidth=\wd0
\catcode`?=\active
\def?{\kern\digitwidth}
\setbox0=\hbox{\rm--}
\minuswidth=\wd0
\catcode`!=\active
\def!{\kern\minuswidth}
\setbox0=\hbox{\rm:}
\colonwidth=\wd0
\catcode`<=\active
\def<{\kern\colonwidth}
{ \tabcolsep=0.3cm
\renewcommand{\baselinestretch}{1.0} \small\normalsize % force adjust
\begin{tabular}{llcccr}
\hline
\hline
\rule{0cm}{0.5cm}
Cepheid & ???!$\gamma$ & $\sigma_{fit}$ & $\sigma_{MC}$ & $\chi^2_{\nu}$ &
Period$^a$\\[2mm]
\hline
 11447-6153     & ?!29.1<       & 2.0  & 1.9    & 1.5  & ?6.4282 \\
 11465-6209     & ??!0.9:       & 4.0  & 3.8    & 4.4  & 11.0984 \\
 11492-6257     & ??!6.7<       & 1.6  & 2.5    & 0.5  & ?3.6798 \\
 11521-6200     & ?!20.4<       & 1.7  & 1.7    & 1.2  & ?6.6039 \\
 12003-6213     & ??!2.3<       & 1.8  & 2.2    & 1.1  & ?9.1266 \\
 13190-6235     & ?--31.7<       & 2.2  & 3.2    & 1.9  & 10.1576 \\
 13240-6245     & ?--29.9:       & 3.3  & 8.6    & 0.6  & 15.0598 \\
 13323-6224     & ?--48.3<       & 3.0  & 2.1    & 2.2  & ?4.2424 \\
\hline
\hline
\end{tabular}
} \\  % tabcolsep redef
\vspace{2mm}
$^a$ Avruch (1991).

\end{table}

\begin{table}
	\centering
	\mytabcaption{Table~\ref{tab:l300parms}. Calculated Parameters for New Cepheids}\label{tab:l300parms}
	\vspace{2mm}
\newdimen\digitwidth
\newdimen\minuswidth
\newdimen\colonwidth
\setbox0=\hbox{\rm0}
\digitwidth=\wd0
\catcode`?=\active
\def?{\kern\digitwidth}
\setbox0=\hbox{\rm--}
\minuswidth=\wd0
\catcode`!=\active
\def!{\kern\minuswidth}
\setbox0=\hbox{\rm:}
\colonwidth=\wd0
\catcode`<=\active
\def<{\kern\colonwidth}
{ \tabcolsep=0.3cm
\renewcommand{\baselinestretch}{1.0} \small\normalsize % force adjust
\begin{tabular}{lcccccr}
\hline
\hline
\rule{0cm}{0.5cm}
Cepheid & Amp$(V)$ & $\langle K \rangle$ & $ \sigma_{\langle K \rangle} $ &
JD${}_{Vmax}^a $ & JD${}_{Kmax}^a$ & Period \\[2mm]
\hline
 11447-6153     & 0.65  & 10.092  & 0.013  & 7962.97 & 8337.43 & ?6.4282 \\
 11465-6209     & 0.88  & ?8.558  & 0.014  & 7968.7? & 8337.84 & 11.0984 \\
 11492-6257     & 0.62  & 10.649  & 0.009  & 7959.80 & 8336.00 & ?3.6529 \\
 11521-6200     & 0.63  & ?9.645  & 0.014  & 7958.40 & 8335.16 & ?6.5763 \\
 12003-6213     & 0.72  & ?9.125  & 0.012  & 7958.67 & 8339.86 & ?9.0131 \\
 13190-6235     & 0.82  & ?8.739  & 0.015  & 7958.76 & 8342.12 & 10.3001 \\
 13240-6245     & 1.21  & ?7.618  & 0.022  & 7966.94 & 8337.60 & 15.2158 \\
 13323-6224     & 0.46  & ?7.910  & 0.012  & 7961.75 & 8336.63 & ?4.2424 \\
\hline
\hline
\end{tabular}
} \\  % tabcolsep redef
\vspace{2mm}
$^a$ Modulo 2,440,000; $\sigma \simeq 0.025 \times$ period.

\end{table}

\begin{table}[p]
	\centering
	\mytabcaption{Table~\ref{tab:cepdatA}.  Cepheid Data}\label{tab:cepdatA}
	\vspace{2mm}
\newlength{\tcdcola} \newlength{\tcdcolb} \newlength{\tcdcolc} 
\newlength{\tcdcold} \newlength{\tcdcole} \newlength{\tcdcolf} 
\newlength{\tcdcolg} 
\settowidth{\tcdcola}{\tt AQLZZETAX}
\settowidth{\tcdcolb}{\tt 300.00\ }
\settowidth{\tcdcolc}{\tt -13.07\ }
\settowidth{\tcdcold}{\tt -100.0\ }
\settowidth{\tcdcole}{\tt 1.23456\ }
\settowidth{\tcdcolf}{\tt 11.111\ }
\settowidth{\tcdcolg}{\tt 0.123\ }
{ \renewcommand{\baselinestretch}{1.0} \small\normalsize % force adjust
\begin{tabular}{p{4.71in}}
\hline
\hline
\rule{0cm}{0.6cm}
\makebox[1\tcdcola][l]{Cepheid}\ \makebox[1\tcdcolb][c]{$\ell$}\
\makebox[1\tcdcolc]{\em b}\ \makebox[1\tcdcold]{$v_r{}^{a}$}\
\makebox[1\tcdcole]{log {\em P}}\ \makebox[1\tcdcolf]{$\langle V\rangle{}^{b,c}$}\
\makebox[1\tcdcolg]{E(B-V)} \\[2mm]
\hline
\begin{verbatim}
AQL  ETA    40.90  -13.10  -14.8  0.8559   3.897  0.164
AQL  U      30.90  -11.60    1.1  0.8466   6.448  0.389
AQL  SZ     35.60   -2.30   10.5  1.2340   8.617  0.607
AQL  TT     36.00   -3.10    1.9  1.1384   7.147  0.476
AQL  FF     49.20    6.40  -17.4  0.6504   5.372  0.232
AQL  FM     44.30    0.90   -7.0  0.7863   8.268  0.611
AQL  FN     38.50   -3.10   13.1  0.9769   8.382  0.489
AQL  V336   34.20   -2.10   11.5  0.8636   9.875  0.610
AQL  V496   28.20   -7.10    7.9  0.8330   7.720  0.402
AQL  V600   43.90   -2.60    3.1  0.8597  10.034  0.812
ARA  V340  335.20   -3.70  -82.2  1.3183  10.242  0.547
AUR  Y     166.80    4.30    8.5  0.5865   9.614  0.385
AUR  RT    183.10    8.90   21.0  0.5715   5.450  0.076
AUR  RX    165.80   -1.30  -23.3  1.0654   7.664  0.278
AUR  SY    164.70    2.10   -3.5  1.0062   9.081  0.439
CMA  RW    232.00   -3.80   50.0  0.7581  11.146  0.520
CMA  RY    226.00    0.30   32.9  0.6701   8.105  0.253
CMA  RZ    231.20   -1.10   24.6  0.6289   9.698  0.448
CMA  SS    239.20   -4.20   73.1  1.0921   9.939  0.524
CMA  TV    227.20   -2.40   39.0  0.6693  10.566  0.555
CMA  TW    229.20    0.10   66.5  0.8448   9.557  0.351
CMA  VZ    239.90   -4.40   40.1  0.4950   9.371  0.466
CAM  RW    144.90    3.80  -26.5  1.2152   8.657  0.614
CAM  RX    145.90    4.70  -36.2  0.8983   7.685  0.542
CAR  1     283.20   -7.00    3.6  1.5507   3.735  0.183
CAR  U     289.10    0.10    1.7  1.5883   6.281  0.285
CAR  V     275.30  -12.30   13.9  0.8258   7.375  0.187
CAR  Y     285.70   -0.30  -14.5  0.5611   8.102  0.190
CAR  SX    286.70    1.30  -11.4  0.6866   9.100  0.323
CAR  UW    285.60   -1.80  -14.4  0.7280   9.441  0.441
CAR  UX    284.80    0.20    8.1  0.5661   8.266  0.141
CAR  UY    287.20   -3.20    4.0  0.7438   8.928  0.200
CAR  UZ    287.30   -2.30  -20.9  0.7164   9.331  0.198
CAR  WW    288.20    0.00  -13.0  0.6700   9.794  0.388
CAR  WZ    289.20   -1.20  -14.7  1.3619   9.255  0.376
CAR  XX    291.30   -4.90  -10.9  1.1963   9.341  0.344
CAR  XY    291.40   -3.90   -5.6  1.0946   9.334  0.405
CAR  XZ    290.30   -0.80    1.5  1.2214   8.595  0.360
CAR  YZ    285.60   -1.40    1.0  1.2592   8.709  0.386
\end{verbatim} \\
\hline
\end{tabular}
}

\end{table}

\begin{table}[p]
	\centering
	{Table~\ref{tab:cepdatA}---{\em Continued}}\\
	\vspace{2mm}
\settowidth{\tcdcola}{\tt AQLZZETAX}
\settowidth{\tcdcolb}{\tt 300.00\ }
\settowidth{\tcdcolc}{\tt -13.07\ }
\settowidth{\tcdcold}{\tt -100.0\ }
\settowidth{\tcdcole}{\tt 1.23456\ }
\settowidth{\tcdcolf}{\tt 11.111\ }
\settowidth{\tcdcolg}{\tt 0.123\ }
{ \renewcommand{\baselinestretch}{1.0} \small\normalsize % force adjust
\begin{tabular}{p{4.71in}}
\hline
\hline
\rule{0cm}{0.6cm}
\makebox[1\tcdcola][l]{Cepheid}\ \makebox[1\tcdcolb][c]{$\ell$}\
\makebox[1\tcdcolc]{\em b}\ \makebox[1\tcdcold]{$v_r{}^{a}$}\
\makebox[1\tcdcole]{log {\em P}}\ \makebox[1\tcdcolf]{$\langle V\rangle{}^{b,c}$}\
\makebox[1\tcdcolg]{E(B-V)} \\[2mm]
\hline
\begin{verbatim}
CAR  AQ    285.80   -3.30    2.1  0.9899   8.821  0.175
CAR  CC    289.40   -1.60    6.9  0.6776  12.021  0.503
CAR  CF    289.40   -1.60   -3.8  0.7400  12.400  0.602
CAR  CN    283.60   -1.30    9.0  0.6931  10.670  0.407
CAR  CQ    286.20   -1.70   20.6  0.7258  13.400  0.902
CAR  CR    285.70   -0.40   25.3  0.9895  11.566  0.484
CAR  CS    288.60    0.20   10.8  0.8236  12.400  0.702
CAR  CY    289.50   -0.90    8.6  0.6300   9.782  0.380
CAR  DY    288.80   -1.00   -1.3  0.6698  11.394  0.382
CAR  ER    290.10    1.50  -19.0  0.8876   6.807  0.121
CAR  FF    286.90    0.60  -14.5  1.2130  12.300  0.900
CAR  FH    288.40   -1.30   14.0  0.7531  12.500  0.700
CAR  FI    287.80    0.70   13.2  1.1289  11.655  0.683
CAR  FK    288.70   -0.40   29.9  1.3665  12.200  0.900
CAR  FM    289.90   -1.00   39.7  0.8830  12.400  0.800
CAR  FN    289.60   -0.10  -13.0  0.6614  11.623  0.553
CAR  FO    290.50   -2.10  -10.8  1.0152  10.773  0.445
CAR  FQ    290.90   -0.40   -3.1  1.0117  11.821  0.811
CAR  FR    291.10    0.60   -7.3  1.0301   9.675  0.346
CAR  FZ    288.40    0.30   -1.6  0.5536  11.600  0.619
CAR  GI    290.30    2.50  -20.6  0.6465   8.320  0.188
CAR  GS    289.20   -1.80   32.1  0.6080  12.500  0.700
CAR  GZ    284.70   -1.90   -8.5  0.6190  10.239  0.407
CAR  HS    285.30   -1.80    5.9  0.7069  12.300  0.700
CAR  IM    289.10   -0.80    3.8  0.7272  12.300  0.700
CAR  IO    289.30   -0.90   39.3  1.1337  10.938  0.453
CAR  IT    291.50   -1.10  -14.9  0.8770   8.094  0.204
CAS  RS    114.50    0.80  -24.5  0.7991   9.928  0.818
CAS  RW    129.00   -4.60  -71.3  1.1702   9.224  0.408
CAS  RY    115.30   -3.30  -70.5  1.0840   9.957  0.614
CAS  SU    133.50    8.50   -7.0  0.2898   5.970  0.288
CAS  SW    109.70   -1.60  -38.0  0.7357   9.700  0.475
CAS  SY    118.20   -4.10  -47.1  0.6097   9.858  0.448
CAS  SZ    134.80   -1.20  -45.9  1.1344   9.852  0.767
CAS  UZ    125.50   -1.60  -51.0  0.6293  11.351  0.498
CAS  VV    130.40   -2.10  -50.5  0.7930  10.751  0.528
CAS  VW    124.60   -1.10  -58.5  0.7777  10.708  0.458
CAS  XY    122.80   -2.80  -42.0  0.6534   9.979  0.534
CAS  AP    120.90    0.10  -44.5  0.8355  11.528  0.781
\end{verbatim} \\
\hline
\end{tabular}
}

\end{table}

\begin{table}[p]
	\centering
	{Table~\ref{tab:cepdatA}---{\em Continued}}\\
	\vspace{2mm}
\settowidth{\tcdcola}{\tt AQLZZETAX}
\settowidth{\tcdcolb}{\tt 300.00\ }
\settowidth{\tcdcolc}{\tt -13.07\ }
\settowidth{\tcdcold}{\tt -100.0\ }
\settowidth{\tcdcole}{\tt 1.23456\ }
\settowidth{\tcdcolf}{\tt 11.111\ }
\settowidth{\tcdcolg}{\tt 0.123\ }
{ \renewcommand{\baselinestretch}{1.0} \small\normalsize % force adjust
\begin{tabular}{p{4.71in}}
\hline
\hline
\rule{0cm}{0.6cm}
\makebox[1\tcdcola][l]{Cepheid}\ \makebox[1\tcdcolb][c]{$\ell$}\
\makebox[1\tcdcolc]{\em b}\ \makebox[1\tcdcold]{$v_r{}^{a}$}\
\makebox[1\tcdcole]{log {\em P}}\ \makebox[1\tcdcolf]{$\langle V\rangle{}^{b,c}$}\
\makebox[1\tcdcolg]{E(B-V)} \\[2mm]
\hline
\begin{verbatim}
CAS  BP    125.40    2.80  -46.5  0.7974  10.916  0.883
CAS  CD    115.50    1.10  -54.9  0.8922  10.746  0.766
CAS  CF    116.60   -1.10  -77.7  0.6880  11.126  0.539
CAS  CG    116.80   -1.30  -79.3  0.6401  11.355  0.663
CAS  DD    116.80    0.50  -70.1  0.9918   9.878  0.481
CAS  DF    136.00    1.50  -33.0  0.5834  10.856  0.569
CAS  DL    120.30   -2.60  -38.1  0.9031   8.964  0.510
CAS  FM    117.80   -6.20  -29.1  0.7641   9.128  0.346
CAS  V636  127.50    1.10  -24.9  0.9231   7.186  0.786
CEN  V     316.40    3.30  -23.2  0.7399   6.820  0.290
CEN  TX    315.20   -0.60  -52.0  1.2328  10.537  0.982
CEN  UZ    295.00   -0.90  -11.9  0.5230   8.760  0.278
CEN  VW    307.60   -1.60  -30.8  1.1771  10.242  0.433
CEN  XX    309.50    4.60  -18.8  1.0397   7.799  0.264
CEN  AY    292.60    0.40  -15.5  0.7251   8.820  0.309
CEN  AZ    292.80   -0.20  -11.5  0.5066   8.635  0.174
CEN  BB    296.40   -0.70  -15.8  0.6017  10.146  0.386
CEN  BK    298.00   -1.00  -26.3  0.5016  10.063  0.327
CEN  IZ    294.90   -0.50  -19.7  0.7703  12.500  0.800
CEN  KK    294.20    2.70   -2.7  1.0856  11.502  0.608
CEN  KN    307.80   -2.10  -39.7  1.5317   9.855  0.654
CEN  LV    294.30   -1.70  -16.7  0.6968  12.100  0.700
CEN  MY    305.30    1.20  -15.2  0.5704  12.018  1.158
CEN  MZ    305.40   -1.60  -30.6  1.0151  11.546  0.852
CEN  OO    306.90   -0.60  -35.7  1.1099  12.004  1.107
CEN  QY    311.90   -0.20  -58.7  1.2492  11.796  1.398
CEN  V339  313.50   -0.50  -24.4  0.9762   8.689  0.415
CEN  V378  306.10    0.30  -16.5  0.8102   8.464  0.386
CEN  V381  310.80    4.40  -31.8  0.7058   7.659  0.214
CEN  V419  292.10    4.30  -15.2  0.7410   8.181  0.188
CEP  DELTA 105.20    0.50  -17.4  0.7297   3.954  0.113
CEP  AK    105.10    0.40  -45.6  0.8593  11.186  0.664
CEP  CP    100.40    1.10  -40.5  1.2518  10.619  0.644
CEP  CR    107.60    0.30  -32.3  0.7947   9.629  0.712
CIR  AX    315.80    4.00  -25.5  0.7221   5.880  0.167
CRU  R     299.60    1.10  -16.5  0.7654   6.771  0.203
CRU  S     303.30    4.40   -7.1  0.6712   6.599  0.177
CRU  T     299.40    0.40   -9.8  0.8282   6.570  0.204
CRU  X     302.30    3.80  -25.0  0.7938   8.357  0.287
\end{verbatim} \\
\hline
\end{tabular}
}

\end{table}

\begin{table}[p]
	\centering
	{Table~\ref{tab:cepdatA}---{\em Continued}}\\
	\vspace{2mm}
\settowidth{\tcdcola}{\tt AQLZZETAX}
\settowidth{\tcdcolb}{\tt 300.00\ }
\settowidth{\tcdcolc}{\tt -13.07\ }
\settowidth{\tcdcold}{\tt -100.0\ }
\settowidth{\tcdcole}{\tt 1.23456\ }
\settowidth{\tcdcolf}{\tt 11.111\ }
\settowidth{\tcdcolg}{\tt 0.123\ }
{ \renewcommand{\baselinestretch}{1.0} \small\normalsize % force adjust
\begin{tabular}{p{4.71in}}
\hline
\hline
\rule{0cm}{0.6cm}
\makebox[1\tcdcola][l]{Cepheid}\ \makebox[1\tcdcolb][c]{$\ell$}\
\makebox[1\tcdcolc]{\em b}\ \makebox[1\tcdcold]{$v_r{}^{a}$}\
\makebox[1\tcdcole]{log {\em P}}\ \makebox[1\tcdcolf]{$\langle V\rangle{}^{b,c}$}\
\makebox[1\tcdcolg]{E(B-V)} \\[2mm]
\hline
\begin{verbatim}
CRU  SU    299.20   -0.60  -33.6  1.1088   9.822  0.930
CRU  SV    296.80   -0.40  -15.4  0.8453  12.130  0.790
CRU  TY    298.00   -0.20  -14.3  0.6980  11.700  0.800
CRU  VV    299.90   -1.80  -35.1  0.7868  12.300  1.301
CRU  VW    300.90   -0.70   -2.7  0.7214   9.602  0.638
CRU  VX    300.90    1.60  -28.5  1.0868  12.007  0.891
CRU  AD    298.50    0.50  -34.3  0.8060  11.037  0.642
CRU  AG    301.70    3.10   -8.5  0.5840   8.204  0.220
CRU  BG    300.40    3.40  -20.3  0.5241   5.462  0.078
CYG  X      76.90   -4.30    8.3  1.2145   6.396  0.289
CYG  SU     64.80    2.50  -21.4  0.5850   6.857  0.116
CYG  SZ     84.40    4.00  -12.1  1.1792   9.434  0.598
CYG  TX     84.40   -2.30  -20.4  1.1676   9.515  1.093
CYG  VX     82.20   -3.50  -18.0  1.3038  10.073  0.742
CYG  VY     82.90   -4.60  -11.9  0.8953   9.590  0.617
CYG  VZ     91.50   -8.50  -18.5  0.6871   8.957  0.290
CYG  BZ     84.80    1.40  -13.2  1.0061  10.232  0.870
CYG  CD     71.10    1.40  -11.6  1.2323   8.957  0.493
CYG  DT     76.50  -10.80   -1.6  0.3978   5.774  0.065
CYG  GH     66.50   -0.10  -11.1  0.8931   9.937  0.626
CYG  MW     70.90   -0.60  -16.4  0.7749   9.485  0.642
CYG  V386   85.50   -4.90   -5.1  0.7207   9.631  0.836
CYG  V402   74.10    2.30  -12.7  0.6400   9.871  0.405
CYG  V438   77.60    2.30  -10.2  1.0496  10.939  1.200
CYG  V459   90.50    0.70  -20.7  0.8604  10.601  0.748
CYG  V520   87.50    1.60  -22.8  0.6073  10.853  0.752
CYG  V532   89.00   -3.00  -16.2  0.5164   9.090  0.511
CYG  V1334  83.60   -8.00   -5.2  0.5226   5.885  -.037
CYG  V1726  92.51   -1.61  -15.3  0.6269   9.011  0.311
DOR  BETA  271.70  -32.80    7.3  0.9931   3.755  0.070
GEM  ZETA  195.70   11.90    6.8  1.0066   3.918  0.046
GEM  W     197.40    3.40   -0.1  0.8985   6.951  0.285
GEM  RZ    187.70   -0.10   12.0  0.7427  10.016  0.543
GEM  AA    184.60    2.70    9.5  1.0532   9.735  0.327
GEM  AD    193.30    7.60   45.0  0.5784   9.858  0.180
GEM  BB    199.40    2.30   64.3  0.3632  11.364  0.444
GEM  DX    198.10    2.80   16.0  0.4964  10.742  0.436
LAC  V     106.50   -2.60  -25.4  0.6976   8.939  0.350
LAC  X     106.50   -2.50  -25.3  0.7359   8.407  0.356
\end{verbatim} \\
\hline
\end{tabular}
}

\end{table}

\begin{table}[p]
	\centering
	{Table~\ref{tab:cepdatA}---{\em Continued}}\\
	\vspace{2mm}
\settowidth{\tcdcola}{\tt AQLZZETAX}
\settowidth{\tcdcolb}{\tt 300.00\ }
\settowidth{\tcdcolc}{\tt -13.07\ }
\settowidth{\tcdcold}{\tt -100.0\ }
\settowidth{\tcdcole}{\tt 1.23456\ }
\settowidth{\tcdcolf}{\tt 11.111\ }
\settowidth{\tcdcolg}{\tt 0.123\ }
{ \renewcommand{\baselinestretch}{1.0} \small\normalsize % force adjust
\begin{tabular}{p{4.71in}}
\hline
\hline
\rule{0cm}{0.6cm}
\makebox[1\tcdcola][l]{Cepheid}\ \makebox[1\tcdcolb][c]{$\ell$}\
\makebox[1\tcdcolc]{\em b}\ \makebox[1\tcdcold]{$v_r{}^{a}$}\
\makebox[1\tcdcole]{log {\em P}}\ \makebox[1\tcdcolf]{$\langle V\rangle{}^{b,c}$}\
\makebox[1\tcdcolg]{E(B-V)} \\[2mm]
\hline
\begin{verbatim}
LAC  Y      98.70   -4.00  -22.0  0.6359   9.150  0.225
LAC  Z     105.80   -1.60  -30.0  1.0369   8.418  0.394
LAC  RR    105.60   -2.00  -38.8  0.8073   8.844  0.348
LAC  BG     93.00   -9.30  -18.6  0.7269   8.878  0.332
LUP  GH    324.90    3.30  -16.1  0.9678   7.633  0.358
MON  T     203.60   -2.60   22.0  1.4317   6.129  0.218
MON  SV    203.70   -3.70   27.1  1.1828   8.260  0.254
MON  TX    214.10   -0.80   51.0  0.9396  10.965  0.490
MON  TY    213.40    1.20   59.5  0.6055  11.400  0.700
MON  TZ    214.00    1.30   34.0  0.8709  10.775  0.427
MON  XX    215.50   -1.10   66.4  0.7370  11.898  0.566
MON  AA    217.10   -0.30   66.1  0.5953  13.300  0.779
MON  BE    204.90    1.00   45.4  0.4322  10.579  0.590
MON  CV    208.60   -1.80   18.9  0.7309  10.306  0.673
MON  EE    220.00    3.80   68.9  0.6821  12.500  0.469
MON  FG    221.70   -0.10   88.7  0.6529  12.900  1.000
MON  FI    221.50    1.00   85.8  0.5169  12.931  0.515
MON  V465  214.90    3.70   60.7  0.4334  10.380  0.260
MON  V508  208.90    0.90   58.8  0.6164  10.551  0.321
MUS  R     302.10   -6.50    0.1  0.8756   6.285  0.138
MUS  S     299.60   -7.50   -1.7  0.9849   6.123  0.162
MUS  RT    296.50   -5.30   -5.5  0.4894   8.977  0.325
MUS  TZ    296.60   -3.00  -24.6  0.6942  11.759  0.657
MUS  UU    296.80   -3.20  -17.0  1.0658   9.772  0.402
NOR  S     327.80   -5.40    5.8  0.9892   6.430  0.200
NOR  U     325.60   -0.20  -21.8  1.1018   9.249  0.833
NOR  RS    329.10   -1.20  -40.5  0.7923  10.035  0.552
NOR  SY    327.50   -0.70  -32.9  1.1019   9.472  0.745
NOR  TW    330.40    0.30  -56.6  1.0328  11.670  1.234
NOR  GU    330.50   -1.70  -24.5  0.5382  10.425  0.646
NOR  QZ    329.44  -02.12  -38.6  0.7300   8.866  0.278
NOR  V340  329.72  -02.27  -40.0  1.0526   8.381  0.332
OPH  Y      20.60   10.10   -6.6  1.2336   6.167  0.620
OPH  BF      9.90    7.10  -29.0  0.6094   7.331  0.252
ORI  RS    196.60    0.30   40.5  0.8789   8.412  0.380
ORI  CR    195.90   -3.90   22.4  0.6912  12.296  0.538
ORI  CS    198.00   -4.50   15.5  0.5898  11.390  0.392
ORI  GQ    199.80   -4.30   43.9  0.9353   8.965  0.281
PER  SV    162.60   -1.50   -4.5  1.0465   8.989  0.421
\end{verbatim} \\
\hline
\end{tabular}
}

\end{table}

\begin{table}[p]
	\centering
	{Table~\ref{tab:cepdatA}---{\em Continued}}\\
	\vspace{2mm}
\settowidth{\tcdcola}{\tt AQLZZETAX}
\settowidth{\tcdcolb}{\tt 300.00\ }
\settowidth{\tcdcolc}{\tt -13.07\ }
\settowidth{\tcdcold}{\tt -100.0\ }
\settowidth{\tcdcole}{\tt 1.23456\ }
\settowidth{\tcdcolf}{\tt 11.111\ }
\settowidth{\tcdcolg}{\tt 0.123\ }
{ \renewcommand{\baselinestretch}{1.0} \small\normalsize % force adjust
\begin{tabular}{p{4.71in}}
\hline
\hline
\rule{0cm}{0.6cm}
\makebox[1\tcdcola][l]{Cepheid}\ \makebox[1\tcdcolb][c]{$\ell$}\
\makebox[1\tcdcolc]{\em b}\ \makebox[1\tcdcold]{$v_r{}^{a}$}\
\makebox[1\tcdcole]{log {\em P}}\ \makebox[1\tcdcolf]{$\langle V\rangle{}^{b,c}$}\
\makebox[1\tcdcolg]{E(B-V)} \\[2mm]
\hline
\begin{verbatim}
PER  SX    158.90   -6.40    5.5  0.6325  11.166  0.471
PER  UX    133.60   -3.10  -41.5  0.6595  11.608  0.514
PER  UY    135.90   -1.40  -45.0  0.7296  11.339  0.857
PER  VX    132.80   -3.00  -35.4  1.0372   9.305  0.494
PER  VY    135.10   -1.70  -39.5  0.7429  11.255  0.877
PER  AS    154.14   -0.88  -25.5  0.6966   9.726  0.672
PER  AW    166.60   -5.40    6.9  0.8104   7.486  0.511
PER  V440  135.90   -5.20  -26.1  0.8791   6.247  0.276
PUP  X     236.10   -0.80   65.3  1.4143   8.490  0.429
PUP  RS    252.40   -0.20   22.1  1.6169   7.028  0.431
PUP  VW    235.36  -00.62  +24.0  0.6320  11.382  0.493
PUP  VX    237.00   -1.30    8.8  0.4789   8.315  0.152
PUP  VZ    243.40   -3.30   63.3  1.3648   9.595  0.454
PUP  WX    241.20   -1.40   54.6  0.9512   9.058  0.317
PUP  WZ    241.80    3.30   64.0  0.7013  10.301  0.228
PUP  AQ    246.20    0.00   59.5  1.4774   8.755  0.491
PUP  AT    254.30   -1.60   26.7  0.8238   8.001  0.195
PUP  BM    244.50   -1.00   64.4  0.8573  10.811  0.581
PUP  BN    247.90    1.00   65.7  1.1359   9.922  0.424
PUP  HW    244.80    0.80  116.2  1.1288  12.129  0.681
PUP  LS    246.40    0.10   77.4  1.1506  10.363  0.460
PUP  MY    261.30  -12.90   12.7  0.7555   5.666  0.088
SGE  S      55.20   -6.10  -10.1  0.9233   5.622  0.144
SGE  GY     54.90   -0.60   15.6  1.7107  10.208  1.268
SGR  U      13.70   -4.50    2.6  0.8290   6.714  0.393
SGR  W       1.60   -4.00  -28.3  0.8805   4.669  0.130
SGR  X       1.20    0.20  -13.7  0.8458   4.561  0.207
SGR  Y      12.80   -2.10   -2.5  0.7614   5.742  0.214
SGR  VY     10.13  -01.07   -6.0  1.1322  11.529  1.185
SGR  WZ     12.10   -1.30  -15.7  1.3395   8.030  0.450
SGR  XX     15.00   -1.90    2.0  0.8078   8.851  0.519
SGR  YZ     17.80   -7.10   18.5  0.9742   7.350  0.293
SGR  AP      8.10   -2.40  -15.0  0.7040   6.942  0.203
SGR  AV      7.50   -0.60    0.0  1.1878  11.425  1.170
SGR  AY     13.30   -2.40  -26.5  0.8176  10.520  0.857
SGR  BB     14.70   -9.00    7.6  0.8220   6.931  0.286
SGR  V350   13.80   -8.00   13.1  0.7114   7.460  0.311
SGR  V773    2.90   -0.50   -0.9  0.7597  12.387  1.488
SGR  V1954  10.56   -1.73   12.7  0.7909  10.861  0.818
\end{verbatim} \\
\hline
\end{tabular}
}

\end{table}

\begin{table}[p]
	\centering
	{Table~\ref{tab:cepdatA}---{\em Continued}}\\
	\vspace{2mm}
\settowidth{\tcdcola}{\tt AQLZZETAX}
\settowidth{\tcdcolb}{\tt 300.00\ }
\settowidth{\tcdcolc}{\tt -13.07\ }
\settowidth{\tcdcold}{\tt -100.0\ }
\settowidth{\tcdcole}{\tt 1.23456\ }
\settowidth{\tcdcolf}{\tt 11.111\ }
\settowidth{\tcdcolg}{\tt 0.123\ }
{ \renewcommand{\baselinestretch}{1.0} \small\normalsize % force adjust
\begin{tabular}{p{4.71in}}
\hline
\hline
\rule{0cm}{0.6cm}
\makebox[1\tcdcola][l]{Cepheid}\ \makebox[1\tcdcolb][c]{$\ell$}\
\makebox[1\tcdcolc]{\em b}\ \makebox[1\tcdcold]{$v_r{}^{a}$}\
\makebox[1\tcdcole]{log {\em P}}\ \makebox[1\tcdcolf]{$\langle V\rangle{}^{b,c}$}\
\makebox[1\tcdcolg]{E(B-V)} \\[2mm]
\hline
\begin{verbatim}
SCO  RV    350.40    5.70  -16.6  0.7825   7.041  0.338
SCO  RY    356.50   -3.40  -17.7  1.3077   8.016  0.729
SCO  KQ    340.40   -0.70  -22.1  1.4575   9.810  0.836
SCO  V470  349.80    0.30   -2.8  1.2112  10.934  1.633
SCO  V482  354.40    0.20   10.7  0.6559   7.973  0.354
SCO  V500  359.00   -1.40  -10.8  0.9693   8.753  0.569
SCO  V636  343.50   -5.20    8.3  0.8323   6.651  0.225
SCT  X      19.00   -1.60    0.0  0.6230  10.011  0.587
SCT  Y      24.00   -0.90   12.3  1.0146   9.659  0.771
SCT  Z      26.80   -0.80   37.2  1.1106   9.595  0.518
SCT  RU     28.40    0.20   -4.8  1.2944   9.465  0.891
SCT  SS     24.20   -1.80   -8.6  0.5648   8.201  0.333
SCT  TY     28.05  +00.12  +25.5  1.0435  10.791  0.943
SCT  UZ     19.16  -01.49  +38.8  1.1686  11.303  0.994
SCT  BX     28.90   -1.70   -2.4  0.8069  12.231  1.216
SCT  CK     26.30   -0.50   -0.4  0.8701  10.608  0.746
SCT  CM     27.20   -0.40   40.8  0.5930  11.101  0.724
SCT  CN     28.10    0.00   19.7  0.9997  12.470  1.170
SCT  EV     23.50   -0.50   17.5  0.4901  10.136  0.641
SCT  V367   21.60   -0.80   -8.4  0.7989  11.604  1.186
SER  AA     30.80    1.80  -44.3  1.2340  12.234  1.318
TAU  ST    193.10   -8.10    0.5  0.6057   8.199  0.350
TAU  SZ    179.50  -18.70    0.3  0.4982   6.530  0.295
TRA  R     317.00   -7.80  -13.2  0.5301   6.650  0.144
TRA  S     322.10    8.20    3.9  0.8009   6.394  0.120
TRA  U     323.20   -8.00  -13.1  0.4098   7.940  0.109
UMI  ALPHA 123.30   26.50  -17.4  0.5988   1.973  0.024
VEL  T     265.50   -3.80    5.7  0.6665   8.030  0.283
VEL  V     276.60   -4.20  -27.9  0.6406   7.570  0.218
VEL  RZ    262.90   -1.90   24.1  1.3096   7.087  0.332
VEL  ST    268.80   -4.80    7.4  0.7677   9.707  0.483
VEL  SV    286.00    2.40    3.5  1.1491   8.576  0.383
VEL  SW    266.20   -3.00   22.9  1.3706   8.126  0.344
VEL  SX    265.50   -2.20   30.9  0.9800   8.290  0.255
VEL  XX    284.80    2.20   15.0  0.8442  10.662  0.545
VEL  AB    283.00    0.60   28.1  0.7956  13.500  1.301
VEL  AE    276.10   -0.60   14.8  0.8533  10.254  0.630
VEL  AH    262.40   -7.00   24.2  0.4953   5.708  0.097
VEL  AP    263.00   -1.40   26.3  0.4953  10.053  0.494
\end{verbatim} \\
\hline
\end{tabular}
}

\end{table}

\begin{table}[p]
	\centering
	{Table~\ref{tab:cepdatA}---{\em Continued}}\\
	\vspace{2mm}
\settowidth{\tcdcola}{\tt AQLZZETAX}
\settowidth{\tcdcolb}{\tt 300.00\ }
\settowidth{\tcdcolc}{\tt -13.07\ }
\settowidth{\tcdcold}{\tt -100.0\ }
\settowidth{\tcdcole}{\tt 1.23456\ }
\settowidth{\tcdcolf}{\tt 11.111\ }
\settowidth{\tcdcolg}{\tt 0.123\ }
{ \renewcommand{\baselinestretch}{1.0} \small\normalsize % force adjust
\begin{tabular}{p{4.71in}}
\hline
\hline
\rule{0cm}{0.6cm}
\makebox[1\tcdcola][l]{Cepheid}\ \makebox[1\tcdcolb][c]{$\ell$}\
\makebox[1\tcdcolc]{\em b}\ \makebox[1\tcdcold]{$v_r{}^{a}$}\
\makebox[1\tcdcole]{log {\em P}}\ \makebox[1\tcdcolf]{$\langle V\rangle{}^{b,c}$}\
\makebox[1\tcdcolg]{E(B-V)} \\[2mm]
\hline
\begin{verbatim}
VEL  AX    263.30   -7.70   22.1  0.4140   8.200  0.232
VEL  BG    271.90   -2.60    7.9  0.8404   7.662  0.433
VEL  BH    259.80   -1.30   51.6  0.8574  11.600  1.300
VEL  CP    267.60   -3.20   73.9  0.9930  12.756  0.855
VEL  CS    277.09   -0.77   26.8  0.7712  11.688  0.792
VEL  CX    272.40   -3.40   16.0  0.7962  11.370  0.713
VEL  DD    271.50   -1.40   26.0  1.1204  12.536  0.911
VEL  DP    275.40   -1.30   29.8  0.7391  11.825  0.892
VEL  DR    273.20    1.30   20.6  1.0492   9.526  0.647
VEL  EX    274.10   -2.20   33.5  1.1217  11.567  0.763
VEL  EZ    274.93  -01.94  +92.2  1.5383  12.440  1.091
VUL  T      72.10  -10.20   -0.9  0.6469   5.749  0.088
VUL  U      56.10   -0.30  -12.1  0.9026   7.127  0.619
VUL  X      63.80   -1.30  -16.1  0.8007   8.848  0.793
VUL  SV     63.90    0.30    1.9  1.6542   7.208  0.543
VUL  DG     64.90   -0.90    2.6  1.1338  11.375  1.167
\end{verbatim} \\
\hline
\hline
\end{tabular} \\
\vspace{2mm}
\raggedright
\makebox[1in]{} ${}^a$ See text for references \\
\makebox[1in]{} ${}^b$ Caldwell \& Coulson (1987) and references therein \\
\makebox[1in]{} ${}^c$ Pont \etal\ (1994) \\
\centering
}

\end{table}

\begin{table}
	\centering
	\mytabcaption{Table~\ref{tab:modelparA}. Axisymmetric Model Parameters}
	\label{tab:modelparA}
	\vspace{2mm}
\newdimen\digitwidth
\newdimen\minuswidth
\newdimen\colonwidth
\setbox0=\hbox{\rm0}
\digitwidth=\wd0
\catcode`?=\active
\def?{\kern\digitwidth}
\setbox0=\hbox{\rm--}
\minuswidth=\wd0
\catcode`!=\active
\def!{\kern\minuswidth}
\setbox0=\hbox{\rm:}
\colonwidth=\wd0
\catcode`<=\active
\def<{\kern\colonwidth}
{ \tabcolsep=0.3cm
\renewcommand{\baselinestretch}{1.0} \small\normalsize % force adjust
\begin{tabular}{ccrrrrrr}
\hline
\hline
\rule{0cm}{0.5cm}
Model & & \multicolumn{1}{c}{$2AR_0$} & \multicolumn{1}{c}{$m_0$} &
     \multicolumn{1}{c}{$u_0$} & \multicolumn{1}{c}{$v_0$} & 
     \multicolumn{1}{c}{$\delta v_r$} & \multicolumn{1}{c}{$R_0$} \\
 & & \multicolumn{1}{c}{km s$^{-1}$} & \multicolumn{1}{c}{mag} &
     \multicolumn{1}{c}{km s$^{-1}$} & \multicolumn{1}{c}{km s$^{-1}$} &
     \multicolumn{1}{c}{km s$^{-1}$} & \multicolumn{1}{c}{kpc} \\[2mm]
\hline
\rule{0cm}{0.5cm}
      A1 & $n=294$ & 241     & 10.45     & -9.5     & 14.1     & 3.2      & 8.13 \\
All Data &         & $\pm13$ & $\pm$0.11 & $\pm$1.3 & $\pm$1.1 & $\pm$0.9 & $\pm$0.42\\
\rule{0cm}{0.8cm}
    A1.1 & $n=288$ & 247     & 10.42     & -8.8     & 13.9     & 2.9      & 8.02 \\
Pruned   &         & $\pm12$ & $\pm$0.09 & $\pm$1.2 & $\pm$1.0 & $\pm$0.8 & $\pm$0.35\\
\rule{0cm}{0.8cm}
    A1.2 & $n=288$ & 253 & 10.35 & -8.4 & 13.7 & 2.9 & 7.76 \\
F90 $E_{B-V}$ &         & $\pm12$ & $\pm$0.09 & $\pm$1.2 & $\pm$1.0 & $\pm$0.8 & $\pm$0.35\\
\rule{0cm}{1.2cm}
      A2   & $n=184$  & 225  & 10.40 & -8.6 & 12.8 & 3.2   & 7.94 \\
CC Data & & $\pm19$ & $\pm$0.19 & $\pm$1.5 & $\pm$1.3 & $\pm$1.0 & $\pm$0.72\\
\rule{0cm}{0.8cm}
    A2.1   & $n=184$  & 236  & 10.40 & -7.0 & 13.2 & 3.1   & \\
New $v_r$  &   & $\pm19$ & $\pm$0.17 & $\pm$1.4 & $\pm$1.2 & $\pm$0.9 & \\
\rule{0cm}{0.8cm}
      A2.2   & $n=184$  & 227  & 10.30 & -8.2 & 12.6 & 3.3   & \\
F90 $E_{B-V}$  &   & $\pm19$ & $\pm$0.17 & $\pm$1.4 & $\pm$1.2 & $\pm$0.9 & \\
\rule{0cm}{0.8cm}
      A2.3   & $n=184$  & 240  & 10.32 & -6.5 & 12.7 & 3.3   & 7.66 \\
Both New  &   & $\pm19$ & $\pm$0.17 & $\pm$1.4 & $\pm$1.2 & $\pm$0.9 & $\pm$0.60\\
\rule{0cm}{1.2cm}
      A3 & $n=266$ & 255     & 10.42     & -9.7     & 13.6     & 0.0      & 8.02 \\
PMB Data &         & $\pm13$ & $\pm$0.12 & $\pm$1.2 & $\pm$1.1 &          & $\pm$0.46\\
\rule{0cm}{0.8cm}
      A3.1 & $n=266$ & 247     & 10.33     & -9.4     & 13.9     & 2.2      & 7.69 \\
PMB w/$\delta v_r$
         &         & $\pm13$ & $\pm$0.11 & $\pm$1.2 & $\pm$1.0 & $\pm$0.8 & $\pm$0.40\\
\hline
\hline
\end{tabular}
} \\

\end{table}

\begin{table}
	\centering
	\mytabcaption{Table~\ref{tab:modelparA1}. Axisymmetric Model Covariances}
	\label{tab:modelparA1}
	\vspace{2mm}
\newdimen\digitwidth
\newdimen\minuswidth
\newdimen\colonwidth
\setbox0=\hbox{\rm0}
\digitwidth=\wd0
\catcode`?=\active
\def?{\kern\digitwidth}
\setbox0=\hbox{\rm--}
\minuswidth=\wd0
\catcode`!=\active
\def!{\kern\minuswidth}
\setbox0=\hbox{\rm:}
\colonwidth=\wd0
\catcode`<=\active
\def<{\kern\colonwidth}
{ \tabcolsep=0.3cm
\renewcommand{\baselinestretch}{1.0} \small\normalsize % force adjust
\begin{tabular}{ccrrrrrr}
\hline
\hline
\rule{0cm}{0.5cm}
 & & \multicolumn{1}{c}{$2AR_0$} & \multicolumn{1}{c}{$m_0$} &
     \multicolumn{1}{c}{$u_0$} & \multicolumn{1}{c}{$v_0$} & 
     \multicolumn{1}{c}{$\delta v_r$} & \multicolumn{1}{c}{$R_0$} \\
 & & \multicolumn{1}{c}{km s$^{-1}$} & \multicolumn{1}{c}{mag} &
     \multicolumn{1}{c}{km s$^{-1}$} & \multicolumn{1}{c}{km s$^{-1}$} &
     \multicolumn{1}{c}{km s$^{-1}$} & \multicolumn{1}{c}{kpc} \\[2mm]
\hline
\rule{0cm}{0.5cm}
Model A1.1 & $n=288$ & 247     & 10.42     & -8.8     & 13.9     & 2.9      & 8.02 \\
Pruned   &         & $\pm12$ & $\pm$0.09 & $\pm$1.2 & $\pm$1.0 & $\pm$0.8 & $\pm$0.35\\
\rule{0cm}{0.4cm}
          & $2AR_0$  & 1.00 &  0.80 & 0.22 & 0.25 & --0.20 & \\
          & $m_0$    &      &  1.00 & 0.18 & 0.38 & --0.23 & \\
          & $u_0$    &      &       & 1.00 & --0.03 &  0.13 & \\
          & $v_0$    &      &       &      & 1.00 &  0.13 & \\
\rule{0cm}{0.8cm}
Model A1.4 & $n=280$ & 253     & 10.47     & -8.7     & 14.1     & 2.8      & 8.20 \\
No New     &         & $\pm14$ & $\pm$0.12 & $\pm$1.2 & $\pm$1.0 & $\pm$0.8 & $\pm$0.47\\
\rule{0cm}{0.4cm}
          & $2AR_0$  & 1.00 &  0.86 & 0.22 & 0.29 & --0.20 & \\
          & $m_0$    &      &  1.00 & 0.18 & 0.39 & --0.22 & \\
          & $u_0$    &      &       & 1.00 & --0.01 &  0.12 & \\
          & $v_0$    &      &       &      & 1.00 &  0.11 & \\
\rule{0cm}{0.8cm}
Model A2   & $n=184$  & 225  & 10.40 & -8.6 & 12.8 & 3.2   & 7.94 \\
CC Data & & $\pm19$ & $\pm$0.19 & $\pm$1.5 & $\pm$1.3 & $\pm$1.0 & $\pm$0.72\\
\rule{0cm}{0.4cm}
          & $2AR_0$  & 1.00 &  0.88 & 0.22 & 0.27 & --0.09 & \\
          & $m_0$    &      &  1.00 & 0.17 & 0.31 & --0.14 & \\
          & $u_0$    &      &       & 1.00 & 0.06 &  0.24 & \\
          & $v_0$    &      &       &      & 1.00 &  0.18 & \\
\hline
\hline
\end{tabular}
} \\

\end{table}

\begin{table}
	\centering
	\mytabcaption{Table~\ref{tab:modelparB}. Non-Axisymmetric Models}
	\label{tab:modelparB}
	\vspace{2mm}
\newdimen\digitwidth
\newdimen\minuswidth
\newdimen\colonwidth
\setbox0=\hbox{\rm0}
\digitwidth=\wd0
\catcode`?=\active
\def?{\kern\digitwidth}
\setbox0=\hbox{\rm--}
\minuswidth=\wd0
\catcode`!=\active
\def!{\kern\minuswidth}
\setbox0=\hbox{\rm:}
\colonwidth=\wd0
\catcode`<=\active
\def<{\kern\colonwidth}
{ \tabcolsep=0.3cm
\renewcommand{\baselinestretch}{1.0} \small\normalsize % force adjust
\begin{tabular}{ccrrrrrr}
\hline
\hline
\rule{0cm}{0.5cm}
Model & & \multicolumn{1}{c}{$v_c$} & \multicolumn{1}{c}{$m_0$} &
     \multicolumn{1}{c}{$u_0$}       & \multicolumn{1}{c}{$v_0$} & 
     \multicolumn{1}{c}{$s(R_0)$}    & \multicolumn{1}{c}{$R_0$} \\
 & & \multicolumn{1}{c}{km s$^{-1}$} & \multicolumn{1}{c}{mag} &
     \multicolumn{1}{c}{km s$^{-1}$} & \multicolumn{1}{c}{km s$^{-1}$} &
     \multicolumn{1}{c}{}            & \multicolumn{1}{c}{kpc} \\[2mm]
\hline
\rule{0cm}{0.5cm}%
      B1 & $n=288$   & 237     & 10.32     & -9.3     & 13.5     & 0.043      & 7.66 \\
Pruned     &         & $\pm12$ & $\pm$0.09 & $\pm$1.1 & $\pm$1.0 & $\pm$0.016 & $\pm$0.32\\
\rule{0cm}{0.5cm}%
    B1.1      & $n=288$ & 242     & 10.26     & -8.8     & 13.3     & 0.044      & 7.45 \\
F90 $E_{B-V}$ &         & $\pm12$ & $\pm$0.09 & $\pm$1.1 & $\pm$1.0 & $\pm$0.016 & $\pm$0.32\\
\rule{0cm}{0.8cm}%
    B2.1 & $R_{v,0}=2.97$   & 235     & 10.37     & -9.3     & 13.5     & 0.044 & 7.83 \\
    B2.2 & $R_{v,0}=3.17$   & 239     & 10.30     & -9.1     & 13.5     & 0.043 & 7.59 \\
\rule{0cm}{0.8cm}%
    B3.1 & $p=-1.0$   & 235     & 10.32     & -8.7     & 13.6     & 0.046 & \\
    B3.2 & $p=-0.5$   & 235     & 10.31     & -9.0     & 13.6     & 0.047 & \\
    B3.3 & $p=+0.5$   & 246     & 10.40     & -9.3     & 13.2     & 0.021 & \\
    B3.4 & $p=+1.0$   & 257     & 10.51     & -9.1     & 13.0     & 0.004 & \\
\rule{0cm}{0.8cm}%
    B3.5 & $\alpha=-0.2$   & 197     & 10.32     & -9.4     & 14.1     & 0.071 & \\
    B3.6 & $\alpha=-0.1$   & 216     & 10.33     & -9.3     & 13.8     & 0.055 & \\
    B3.7 & $\alpha=+0.1$   & 262     & 10.32     & -9.1     & 13.2     & 0.034 & \\
    B3.8 & $\alpha=+0.2$   & 293     & 10.30     & -9.0     & 12.9     & 0.027 & \\
\hline
\hline
\end{tabular}
} \\

\end{table}

\clearpage

\renewcommand{\bottomfraction}{0.0}
\renewcommand{\floatpagefraction}{0.1}
\setcounter{topnumber}{1}	% One per page
\setcounter{bottomnumber}{1}	% One per page
\setcounter{totalnumber}{1}	% One per page
\clearpage

\input{figs}

\end{document}

%% file: figs.tex
\begin{figure}
	\vspace{3.75in}
	\includegraphics{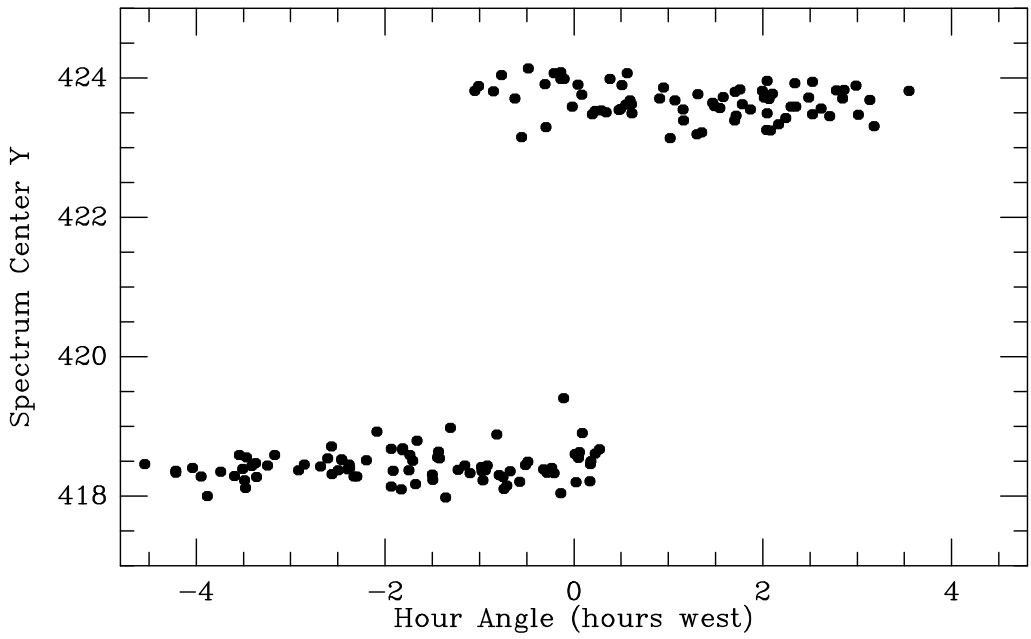}
	\caption[Modspec Flop]{
		The pixel location of the center of the spectrum, along
		the cross-dispersed direction, plotted as a function of
		telescope hour angle.  A significant shift is seen near the
		meridian (with some hysteresis), possibly caused by motion
		of the grism.
	}\label{fig:msflop}
\end{figure}

\begin{figure}
	\vspace{3.75in}
	\includegraphics{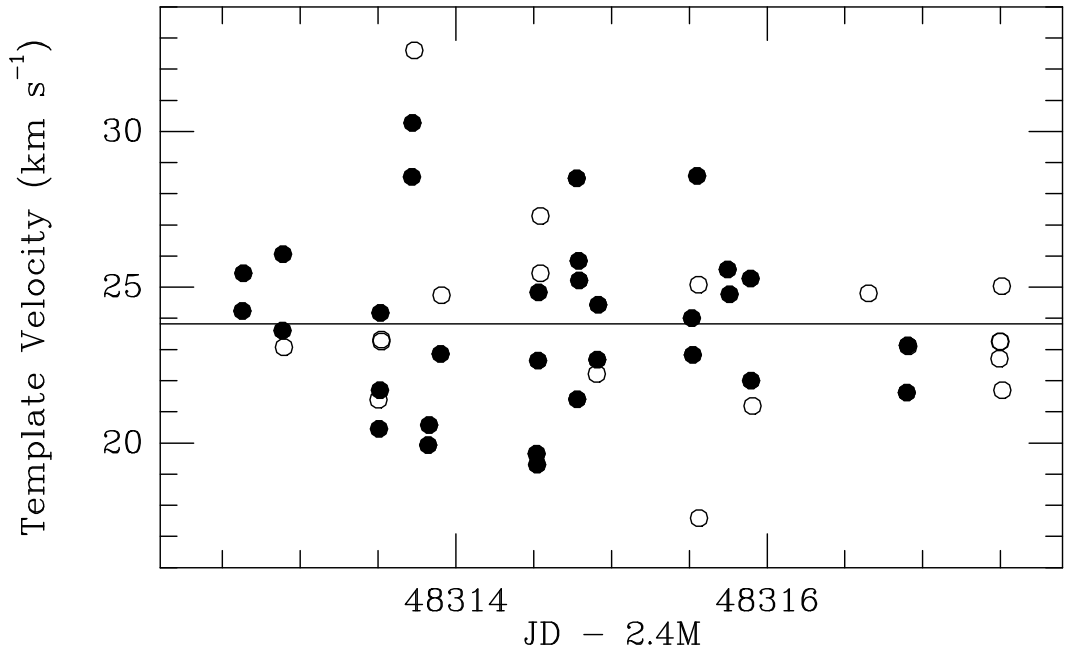}
	\caption[Template Velocities for Feb 91 Run]{
		Velocities of the template calculated from spectra of radial velocity
		standards.  The open and filled points correspond to
		velocities measured from the two different locations of the
		spectrum on the chip; the error of each measurement is
		typically 2.7 \kms.  The solid line shows the adopted mean
		velocity for the template.
	}\label{fig:mstemplvel}
\end{figure}

\begin{figure}
	\vspace{7.5in}
	\includegraphics{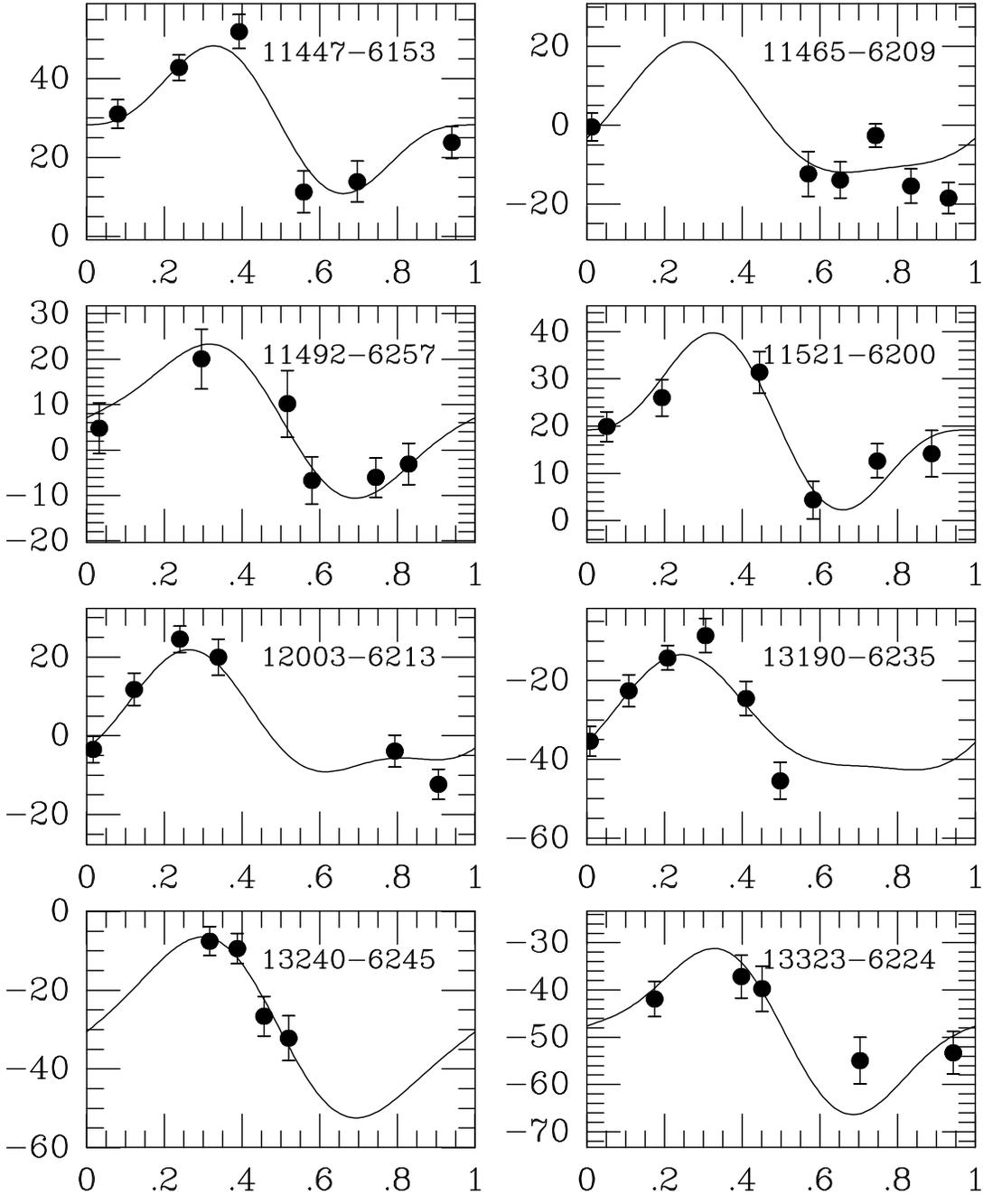}
	\caption[New Cepheid Radial Velocity Curves]{
		Radial velocity curves for the newly discovered Cepheids,
		using the designations of Caldwell \etal.  The curve shape was determined
		from the period, and its position in phase and $\gamma$
		velocity were fit to the radial velocities shown.
	}\label{fig:l300gam}
\end{figure}

\begin{figure}
	\vspace{7.5in}
	\includegraphics{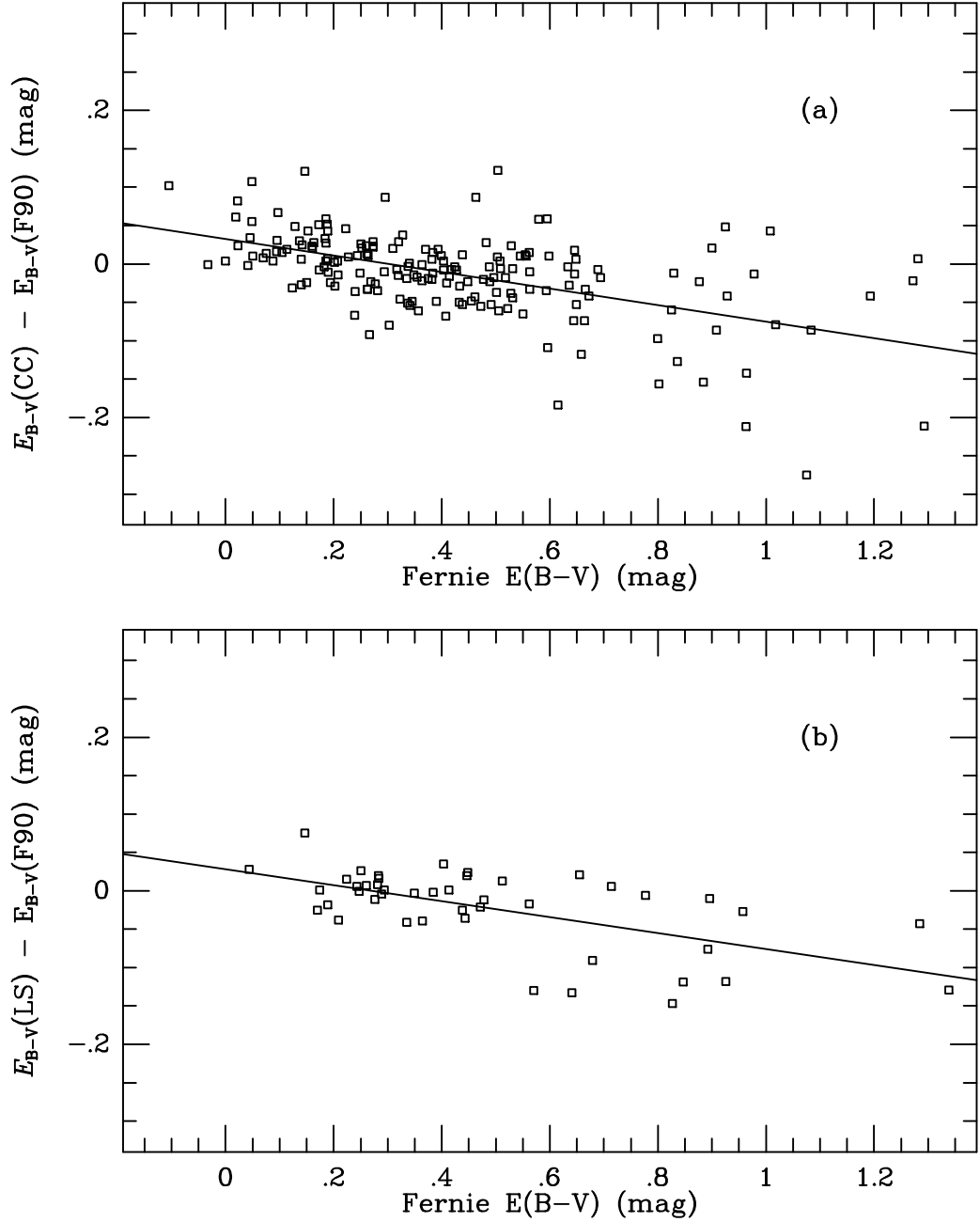}
	\caption[E(B--V) Comparison]{
		A comparison of the reddenings from Fernie (1990)
		and (a) Caldwell \& Coulson (1987); (b) Laney \& Stobie 1994.
		A linear fit to each is shown, revealing a reddening scale
		difference in both.
	}\label{fig:newred}
\end{figure}

\begin{figure}[t]
	\vspace{3.75in}
	\includegraphics{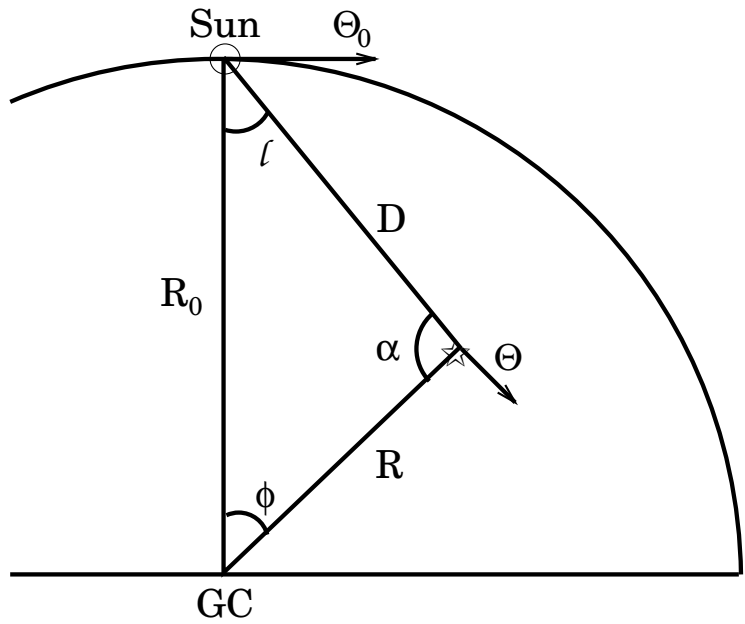}
	\caption[Rotation Curve Schematic]{
		A schematic of Milky Way rotation, with labels
		indicating quantities discussed in the text.  The sun is
		indicated with a circle near the top of the figure, and a
		fiducial Cepheid is indicated with a star.  The Galactic
		center is labeled GC.
	}\label{fig:rotschem}
\end{figure}

\begin{figure}
	\vspace{3.75in}
	\includegraphics{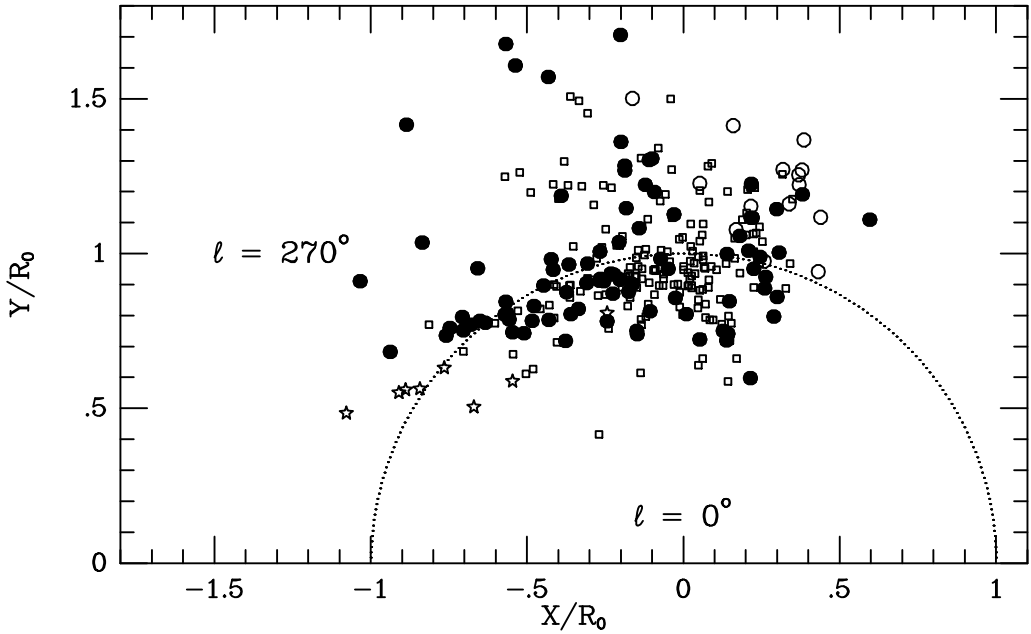}
	\caption[Map of Milky Way Cepheids]{
		Locations of Milky Way Cepheids used in the rotation curve
		models.  Squares indicate Cepheids modeled
		by CC.  Open circles indicate additional Cepheids with new
		reddenings, filled circles Cepheids with new radial
		velocities from MCS and PMB, and stars indicate
		Cepheids newly discovered by Caldwell \etal.
		Cartesian coordinates are shown in units of $R_0$ with the
		Galactic center at (0,0) and the Sun at (0,1).
		The solar circle ($r=1$) is shown with a dotted line.  
	}\label{fig:cepmap}
\end{figure}

\begin{figure}
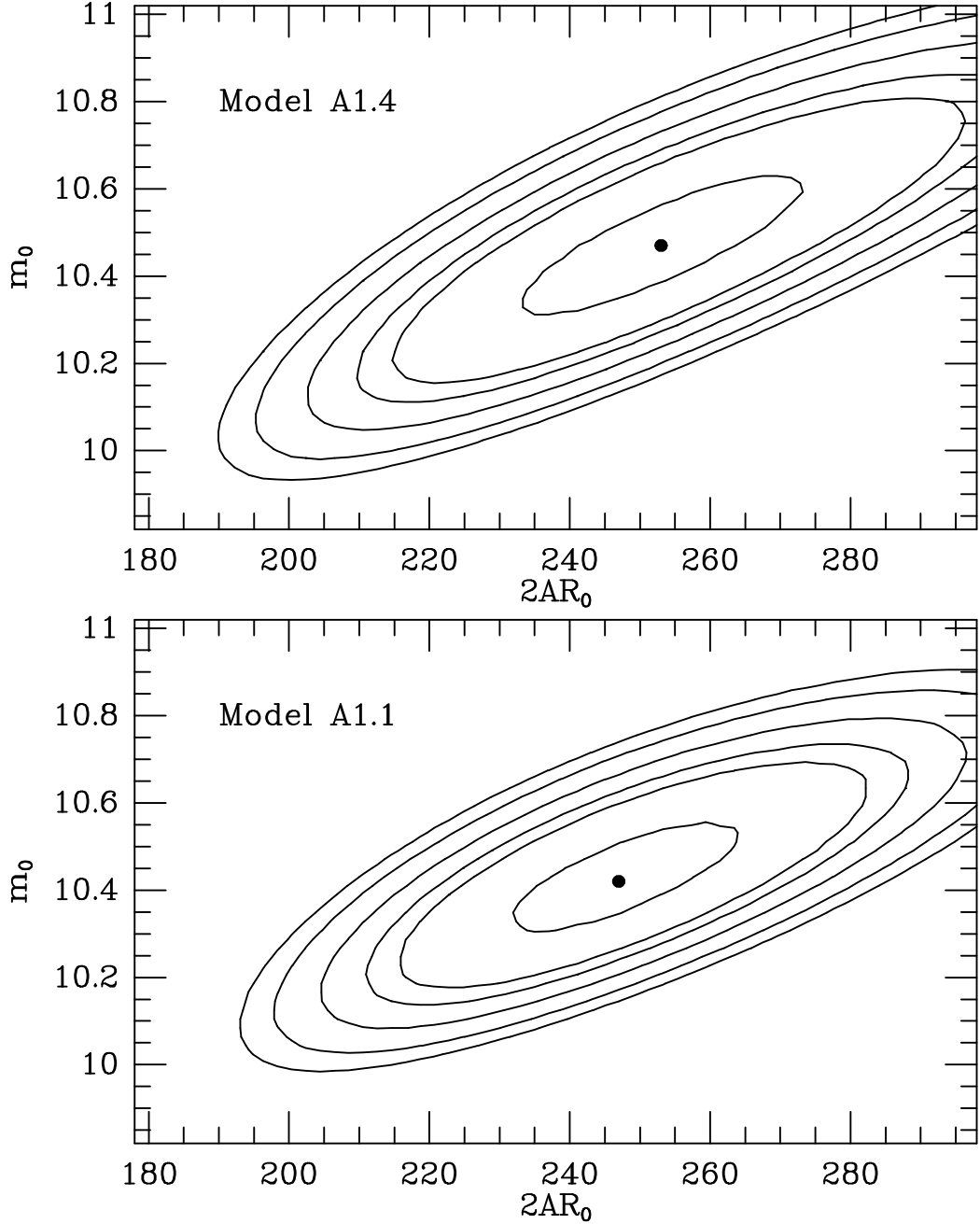

	\vspace{7.5in}
	\includegraphics{psfigs/figure7a.ps}
	\includegraphics{psfigs/figure7b.ps}
	\caption[Model $\chi^2$ Contours]{
		Constant $\delta\chi^2$ contours for two models projected in the
		$2AR_0$, $m_0$ plane.
		Contours shown are (from inside to outside) 1-$\sigma$, 90\%,
		2-$\sigma$,
		99\%, 99.9\%, and 99.99\%.  The top panel is a model using
		all Cepheids with $V$-band data.  The bottom panel is a model
		that also includes CKS Cepheids, where distances were
		determined from $K$-band data.
	}\label{fig:chicontour}
\end{figure}

\begin{figure}
	\vspace{3.5in}
	\includegraphics{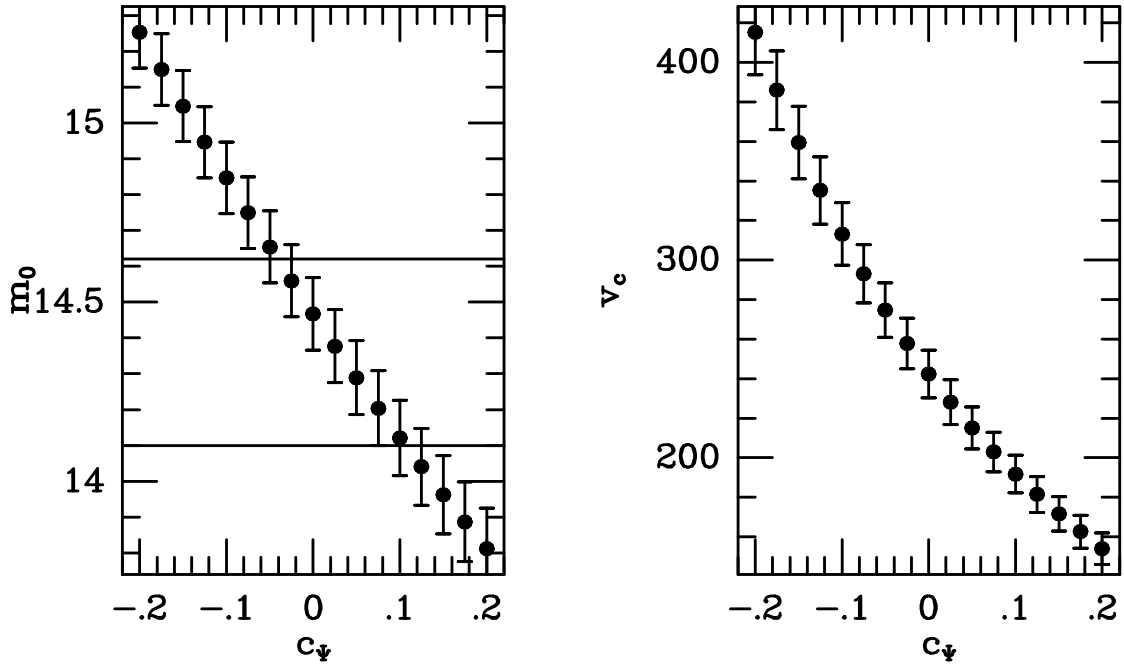}
	\caption[Effects of Ellipticity on $v_c$ and $m_0$]{
		A plot of the best-fit model parameters as a function of
		the symmetric ellipticity component $c(R_0)$.
		The left panel shows the fit distance modulus of the galactic
		center, $(m-M)_0$, the right panel circular
		velocity.  A flat rotation curve and constant ellipticity are assumed.  The
		two horizontal lines represent 1-$\sigma$ errors on other,
		non-kinematic distance measurements (Reid 1993).
	}\label{fig:kt1}
\end{figure}

\begin{figure}
	\vspace{4.0in}
%	\special{psfile=figs/resid/resid-vel-b1.ps hscale=1.36 vscale=1.065
%		hoffset=62
%                 voffset=39}
%	\special{psfile=figs/resid/resid-b1v-cont.ps hscale=0.75 vscale=0.75
%		hoffset=-10
%                 voffset=-40}
	\includegraphics{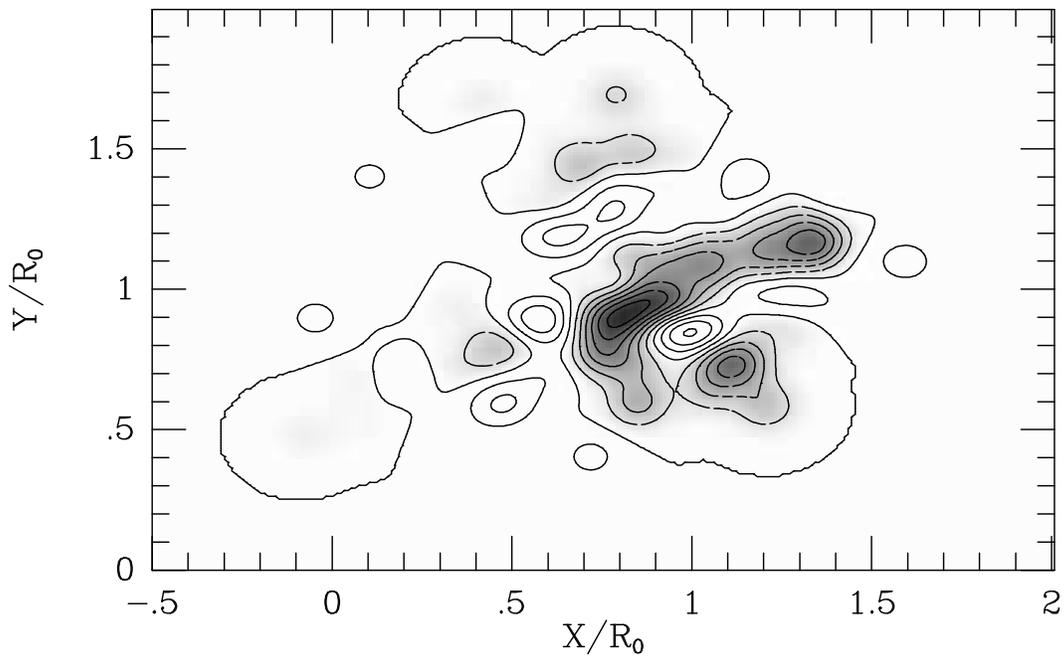}
	\caption[Model residuals]{
		Residual velocities of the stars in Model B1, smoothed with
		a 0.5 kpc gaussian filter.  Contour levels are at
		2 \kms\ increments.  Shaded contours are negative velocity 
		residuals, i.e. the mean motion of stars in these areas is
		toward the sun with respect to the model.
	}\label{fig:b1resid}
\end{figure}

\begin{figure}
	\vspace{3.5in}
	\includegraphics{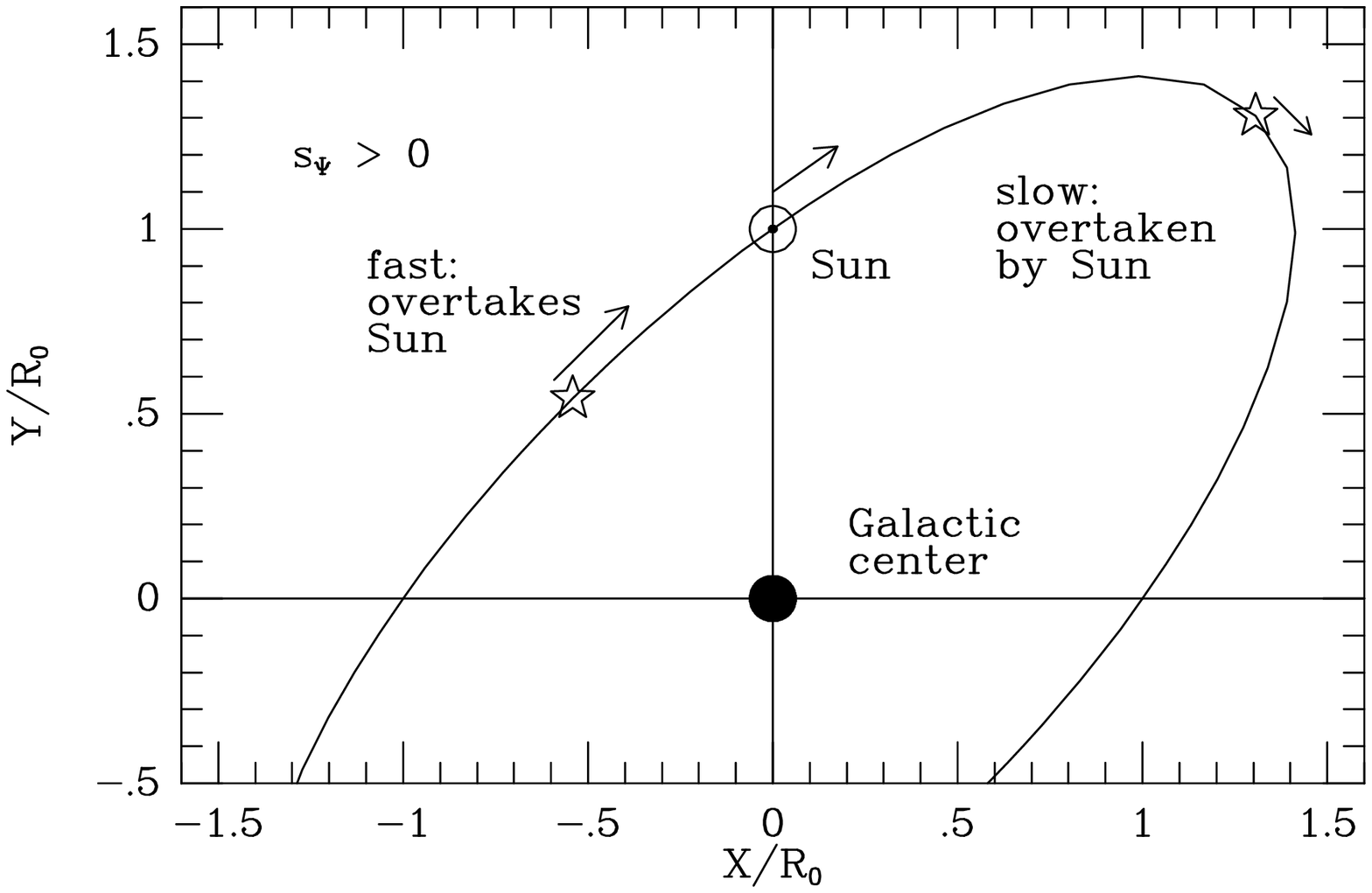}
	\caption[Schematic of Local Ellipticity]{
		A schematic of the closed elliptical orbits that pass
		through the sun in our fiducial model.  The general
		orientation of the orbits is depicted; the ellipticity has
		been exaggerated for clarity.
	}\label{fig:ellipschem}
\end{figure}

%% file: ceph-aph.bbl
\begin{thebibliography}{}

\bibitem[Ajhar \etal\ 1996]{ajh96}
\ajref{Ajhar, E.~A., Grillmair, C.~J., Lauer, T.~R., Baum, W.~A., Faber,
S.~M., Holtzman, J.~A., Lynds, C.~R., \& O'Neil, E.~J.}{1996}{111}{1110}

\bibitem[Alcock \etal\ 1995]{alc95}
\ajref{Alcock, C., Allsman, R.~A., Axelrod, T.~S., Bennett, D.~P., Cook, K.~H.,
Freeman, K.~C., Griest, K., Marshall, S.~L., Peterson, B.~A., \& Pratt,
M.~R.}{1995}{109}{1653}

\bibitem[Alves \etal\ 1995]{alv95}
\journref{Alves, D.~R. \etal}{1995}{\baas}{187}{102.03}

\bibitem[Avruch 1991]{avr91}
Avruch, I.~M.\ 1991, M.S. Thesis, Massachusetts Institute of Technology.

\bibitem[Backer \& Sramek 1987]{bs87}
Backer, D.~C., \& Sramek, R.~A. 1987, in The Galactic Center,
AIP Conf. Proc 155, ed. D. Backer (New York: AIP), p. 163

\bibitem[Bevington 1989]{bev89}
Bevington, P.~R. 1969, {\em Data Reduction and Error Analysis for the
Physical Sciences} (New York: McGraw-Hill)


\bibitem[Binney \etal\ 1991]{bin91}
\mnrasref{Binney, J., Gerhard, O.~E., Stark, A.~A., Bally, J., \& Uchida,
K.~I.}{1991}{252}{210}

\bibitem[Blitz \& Spergel 1991a]{bs91a}
\apjref{Blitz, L., \& Spergel, D.~N.}{1991a}{370}{205}

\bibitem[Blitz \& Spergel 1991b]{bs91b}
\apjref{Blitz, L., \& Spergel, D.~N.}{1991b}{379}{631}

\bibitem[Caldwell \& Coulson 1987]{cc87}
\ajref{Caldwell, J.~A.~R., \& Coulson, I.~M.}{1987}{93}{1090} (CC)

\bibitem[Caldwell, Keane, \& Schechter 1991]{cks91}
\ajref{Caldwell, J.~A.~R., Keane, M.~J., \& Schechter, P.~L.}
{1991}{101}{1763} (CKS)

\bibitem[Caldwell \etal\ 1992]{camsk92}
Caldwell, J., Avruch, I., Metzger, M., 
	Schechter, P., \& Keane, M. 1992, 
	in Variable Stars and Galaxies,
	ed. B. Warner, ASP Conf. Ser. 30, 111.

\bibitem[Carney, Storm, \& Jones 1992]{csj92}
\apjref{Carney, B.~W., Storm, J., \& Jones, R.~V.}{1992}{386}{663} (CSJ)

\bibitem[Carney \etal\ 1995]{car95}
\ajref{Carney, B.~W., Fulbright, J.~P., Terndrup, D.~M., Suntzeff, N.~B., \&
Walker, A.~R.}{1995}{110}{1674}

\bibitem[Clayton, Cardelli, \& Mathis 1989]{ccm89}
\apjref{Clayton, J.~A., Cardelli, G.~C., \& Mathis, J.~S.}{1989}{345}{245}

\bibitem[Cohen \etal\ 1981]{cfpe91}
\apjref{Cohen, J.~G., Frogel, J.~A., Persson, S.~E., \& Elias,
        J.~H.}{1981}{249}{481}


\bibitem[Coulson \& Caldwell 1985]{cc85}
\mnrasref{Coulson, I.~M., \& Caldwell, J.~A.~R.}{1985}{216}{671}

\bibitem[Dean, Warren, \& Cousins 1978]{dwc78}
\mnrasref{Dean, J.~F., Warren, P.~R., \& Cousins, A.~W.~J.}{1978}{183}{569}

\bibitem[Delhaye 1965]{del65}
Delhaye, J. 1965, in Galactic Structure, eds. A. Blaauw \& M.
Schmidt (Chicago: U. of Chicago), p. 61

\bibitem[Dwek \etal\ 1995]{dwek95}
\apjref{
Dwek, E., Arendt, R.~G., Hauser, M.~G., Kelsall, T., Lisse, C.~M., Moseley,
S.~H., Silverberg, R.~F., Sodroski, T.~J., \& Weiland, J.~L.}{1995}{445}{716}

\bibitem[Elias \etal\ 1982]{eli82}
\ajref{Elias, J.~H., \etal.}{1982}{87}{1029}

\bibitem[Evans 1991]{eva91}
\apjref{Evans, N.~R.}{1991}{372}{597}

\bibitem[Feast 1967]{fea67}
\mnrasref{Feast, M.~W.}{1967}{136}{141}

\bibitem[Feast \& Walker 1987]{fw87}
\journref{Feast, M.~W., \& Walker, A.~R.}{1987}{ARA\&A}{25}{345}

\bibitem[Fernie 1987]{fer87}
\ajref{Fernie, J.~D.}{1987}{94}{1003}

\bibitem[Fernie 1990]{fer90}
\apjsref{Fernie, J.~D.}{1990}{72}{153}

\bibitem[Fernie 1994]{fer94}
\apjref{Fernie, J.~D.}{1994}{429}{824}

\bibitem[Fernie \etal\ 1995]{fer95}
Fernie, J.~D., Beattie, B., Evans, N.~R., \& Seager, S. 1995, IBVS No. 4148

\bibitem[Fernie \& McGonegal 1983]{fm83}
\apjref{Fernie, J.~D., \& McGonegal, R.}{1983}{275}{732}

\bibitem[Gieren, Barnes, \& Moffett 1989]{gie89}
\apjref{Gieren, W.~P., Barnes, T.~G., \& Moffett, T.~J.}{1989}{342}{467}

\bibitem[Gieren \& Fouqu\'e 1993]{gf93}
\ajref{Gieren, W.~P., \& Fouqu\'e, P.}{1993}{106}{734}

\bibitem[Gould 1994]{gou94}
\apjref{Gould, A.}{1994}{426}{542}

\bibitem[Harris 1985]{har85}
\ajref{Harris, H.~C.}{1985b}{90}{756}

\bibitem[Iben \& Renzini 1984]{ir84}
\journref{Iben, I., \& Renzini, A.}{1984}{Phys. Rept.}{105}{329}

\bibitem[Joy 1939]{joy39}
\apjref{Joy, A.~H.}{1939}{89}{356}

\bibitem[Kraft \& Schmidt 1963]{ks63}
\apjref{Kraft, R.~P., \& Schmidt, M.}{1963}{137}{249}

\bibitem[Kuijken 1992]{kui92}
\paspref{Kuijken, K.}{1992}{104}{809}

\bibitem[Kuijken \& Tremaine 1994]{kt94}
\apjref{Kuijken, K., \& Tremaine, S.}{1994}{421}{178} (KT)

\bibitem[Laney \& Stobie 1993a]{ls93a}
\mnrasref{Laney, C.~D., \& Stobie, R.~S.}{1993a}{260}{408}

\bibitem[Laney \& Stobie 1993b]{ls93b}
\mnrasref{Laney, C.~D., \& Stobie, R.~S.}{1993b}{263}{921}

\bibitem[Laney \& Stobie 1994]{ls94}
\mnrasref{Laney, C.~D., \& Stobie, R.~S.}{1994}{266}{441}

\bibitem[LeDell 1993]{led93}
LeDell, A.~M.\ 1993, B.S. Thesis, Massachusetts Institute of Technology.

\bibitem[Lewis \& Freeman 1989]{lf89}
\ajref{Lewis, J.~R., \& Freeman, K.~C.}{1989}{97}{139}

\bibitem[McGonegal \etal\ 1983]{mcg83}
\apjref{McGonegal, R., McAlary, C.~W., McLaren, R.~A., \& Madore,
B.~F.}{1983}{269}{641}

\bibitem[Madore \& Freedman 1991]{mf91}
\paspref{Madore, B.~F., \& Freedman, W.~L.}{1991}{103}{933}

\bibitem[Mateo \& Schech\-ter 1989]{doph89}
Mateo, M., \& Schechter, P.~L. 1989, in 1$^{\rm st}$ ESO/ST-ECF Data Analysis
        Workshop, eds. P.~J.~Grosb\o l, F.~Murtagh, \&
        R.~H.~Warmels, p. 69

\bibitem[Maurice \etal\ 1984]{mau84}
\journref{Maurice, E., Mayor, M., Andersen, J., Ardeberg, A., Benz, W.,
Lindgren, H., Imbert, M., Martin, N., Nordstr\"om, B., \& Pr\'evot,
L.}{1984}{A\&AS}{57}{275}

\bibitem[Mayor 1985]{may85}
Mayor, M. 1985, in Stellar Radial Velocities, eds. A.~G.~D.
  Phillip \& D.~W. Latham, p. 299


\bibitem[Mermilliod \etal\ 1987]{mmb87}
\journref{Mermilliod, J.~C., Mayor, M., \& Burki, G.}{1987}{\aap}{70}{389}

\bibitem[Metzger \etal\ 1991]{mcms91}
\apjsref{Metzger, M.~R., Caldwell, J.~A.~R., McCarthy, J.~K,, \& Schechter,
        P.~L.}{1991}{76}{803}

\bibitem[Metzger, Caldwell, \& Schechter 1992]{mcs92}
\ajref{Metzger, M.~R., Caldwell, J.~A.~R., \& Schechter,
P.~L.}{1992}{103}{529} (MCS)

\bibitem[Metzger \& Schechter 1994]{ms94}
\apjref{Metzger, M.~R., \& Schechter, P.~L.}{1994}{420}{177}

\bibitem[Metzger 1994]{met94}
Metzger, M.~R.\ 1994, Ph.D. Thesis, Massachusetts Institute of Technology.

\bibitem[Mihalas \& Binney 1981]{mb81}
Mihalas, D., \& Binney, J. 1981, {\em Galactic Astronomy}
(New York: Freeman)

\bibitem[Moffett \& Barnes 1985]{mb85}
\apjsref{Moffett, T.~J., \& Barnes, T.~G.}{1985}{58}{843}

\bibitem[Moffett \& Barnes 1986]{mb86}
\mnrasref{Moffett, T.~J., \& Barnes, T.~G.}{1986}{219}{45P}

\bibitem[Moffett \& Barnes 1987]{mb87}
\paspref{Moffett, T.~J., \& Barnes, T.~G.}{1987}{99}{1206}

\bibitem[Olson 1975]{ols75}
\paspref{Olson, B.~I.}{1975}{87}{349}

\bibitem[Pont, Mayor, \& Burki 1994a]{pmb94}
\journref{Pont, F., Mayor, M. \& Burki, G.}{1994}
{\aap}{285}{415} (PMB)

\bibitem[Reid \etal\ 1988]{reid88}
\apjref{Reid, M.~J., Schneps, M.~H., Moran, J.~M., Gwinn, C.~R., Genzel, R.,
 Downes, D., \& Roennaeng, B.}{1988}{330}{809}

\bibitem[Reid 1993]{reid93}
\journref{Reid, M.~J.}{1993}{ARA\&A}{31}{345}

\bibitem[Rieke \& Lebofsky 1985]{rie85}
\apjref{Rieke, G.~H., \& Lebofsky, M.~J.}{1985}{288}{618}

\bibitem[Sargent \etal\ 1977]{ssbs77}
\apjref{Sargent, W.~L.~W., Schechter, P.~L., Boksenberg, A., \& Shortridge, K.}
{1977}{212}{326}

\bibitem[Savage \& Mathis 1979]{sm79}
\journref{Savage, B.~D., \& Mathis, J.~D.}{1979}{ARA\&A}{17}{73}

\bibitem[Schechter \etal\ 1992]{sack92}
\ajref{Schechter, P.~L., Avruch, I.~M., Caldwell, J.~A.~R., \& Keane,
M.~J.}{1992}{104}{1930}

\bibitem[Stibbs 1956]{sti56}
\mnrasref{Stibbs, D.~W.~N.}{1956}{116}{453}

\bibitem[Stothers 1988]{sto88}
\apjref{Stothers, R.~B.}{1988}{329}{712}

\bibitem[Turner 1976]{tur76}
\ajref{Turner, D.~G.}{1976}{81}{1125}

\bibitem[Turner 1985]{tur85}
Turner, D.~G. 1985, in Cepheids: Theory and Observations, ed. B.~F.
Madore (Cambridge: Cambridge), p. 209.

\bibitem[van den Bergh 1995]{vdb95}
\apjref{van den Bergh, S.}{1995}{446}{39}

\bibitem[Walker \& Terndrup 1991]{wlt91}
\apjref{Walker, A.~R., \& Terndrup, D.~M.}{1991}{378}{119}

\bibitem[Walker 1992]{wal92}
\apjref{Walker, A.~R.}{1992}{390}{L81}

\bibitem[Weinberg 1992]{wein92}
\apjref{Weinberg, M.}{1992}{384}{81}

\bibitem[Weinberg 1994]{wein94}
\apjref{Weinberg, M.}{1994}{420}{597}

\bibitem[Welch \etal\ 1984]{wel84}
\apjsref{Welch, D.~L., Wieland, F., McAlary, C.~W., McGonegal, R., Madore,
B.~F., McLaren, R.~A., \& Neugebauer, G.}{1984}{54}{547}


\end{thebibliography}
